\expandafter\def\csname ver@fixltx2e.sty\endcsname{}
\documentclass[a4paper,oneside,preprint,12p]{elsarticle}

\usepackage{url}
\usepackage{microtype}

\usepackage{dblfloatfix}
\usepackage[utf8]{inputenc}
\usepackage{natbib}
\usepackage{amsmath}
\usepackage{amsthm}
\usepackage{amssymb}
\usepackage{amsfonts}
\usepackage{algorithmic}
\usepackage{bm}
\usepackage{siunitx}
\usepackage{epstopdf}          
\usepackage{flushend}
\usepackage{parskip}
\usepackage{hyperref}
\usepackage[dvipsnames,table]{xcolor}

\definecolor{bluePolimi}{RGB}{22, 44, 80}
\definecolor{lightBluePolimi}{RGB}{91, 122, 172}
\definecolor{lightlightBluePolimi}{RGB}{140, 160, 190}
\definecolor{greenPolimi}{RGB}{0, 110, 0}
\definecolor{purplePolimi}{RGB}{106, 13, 173}
\definecolor{orangePolimi}{RGB}{255, 128, 13}
\definecolor{redPolimi}{RGB}{160, 0, 0}
\definecolor{yellowPolimi}{RGB}{200, 200, 0}
\definecolor{blackPolimi}{RGB}{36, 36, 36}
\definecolor{greyUD}{RGB}{22, 44, 80}
\definecolor{lightBlueUD}{RGB}{0, 70, 255}
\definecolor{greenUD}{RGB}{0, 110, 0}
\definecolor{redUD}{RGB}{255, 0, 0}

\makeatletter
\hypersetup{colorlinks=true}
\AtBeginDocument{\@ifpackageloaded{hyperref}
  {\def\@linkcolor{blue}
   \def\@anchorcolor{red}
   \def\@citecolor{red}
   \def\@filecolor{red}
   \def\@urlcolor{redPolimi}
   \def\@menucolor{red}
   \def\@pagecolor{cyan}
\begingroup
  \@makeother\`%
  \@makeother\=%
  \edef\x{%
    \edef\noexpand\x{%
      \endgroup
      \noexpand\toks@{%
        \catcode 96=\noexpand\the\catcode`\noexpand\`\relax
        \catcode 61=\noexpand\the\catcode`\noexpand\=\relax
      }%
    }%
    \noexpand\x
  }%
\x
\@makeother\`
\@makeother\=
}{}}
\makeatother

\usepackage[pagewise]{lineno}
\usepackage{framed} 
\usepackage{multicol} 
\usepackage{multirow}
\usepackage{tabularx} 
\usepackage{longtable} 
\usepackage[ruled,vlined]{algorithm2e}

\journal{Nuclear Engineering and Design}
\usepackage{amssymb}

\usepackage[flushright]{rotating}

\usepackage{subcaption} 
\usepackage[printonlyused]{acronym} 
\usepackage{mathtools} 

\usepackage{tikz}
\usetikzlibrary{fit}
\usetikzlibrary{positioning}
\usetikzlibrary{shapes,arrows, arrows.meta} 
\usetikzlibrary{calc} 
\usetikzlibrary{patterns,patterns.meta}

 \newcolumntype{M}[1]{>{\centering\arraybackslash}m{#1}}
 \usepackage{booktabs} 
 \usepackage{comment}

\begin{document}

\newtheorem{theo}{Theorem}
\theoremstyle{definition}
\newtheorem{obs}{Remark}
\newtheorem{Def}{Definition} 

\begin{frontmatter}

\title{Photonuclear treatment for spent fuel radiotoxicity reduction: a case study investigation on minor actinides}

\author[Second,First]{Antonio Cammi$^{*,}$}
\author[Second]{Lorenzo Loi}
\author[Second]{Andrea Missaglia}
\author[Second]{Ludovica Tumminelli}
\author[Second]{Francesca Giacobbo}
\author[Second]{Enrico Padovani}

\cortext[cor1]{Corresponding author. Email address: antonio.cammi@ku.ac.ae}

\address[Second]{Politecnico di Milano, Dept. of Energy, CeSNEF-Nuclear Engineering Division, Nuclear Reactors Group - via La Masa, 34 20156 Milano, Italy}
\address[First]{Emirates Nuclear Technology Center (ENTC), Department of Mechanical and Nuclear Engineering, Khalifa University, Abu Dhabi, 127788, United Arab Emirates}

\begin{abstract}
The management of Spent Nuclear Fuel (SNF) is one of the main challenges in the decommissioning of nuclear power plants. Thermal reactors, such as Light Water Reactors (LWRs), produce significant amounts of minor actinides (MAs) such as Americium, Curium, and Neptunium, which are key contributors to the long-term radiotoxicity and decay heat in SNF. Currently, the long term widely accepted solution is the geological disposal. At the same time, advanced technologies like Partitioning and Transmutation (P\&T) offer promising solutions to reduce SNF long-term radiotoxicity. While most transmutation strategies rely on neutron fluxes, in this study the adoption of photon beam to induce photonuclear reactions in SNF is investigated, without depending on neutron based systems. In particular, the study focuses on the probability of inducing transmutations and fissions on MAs, by leveraging the Giant Dipole Resonance (GDR) region of photonuclear interactions. In the presented case study, the aim was to investigate the effect of a photon driven transmutation of minor actinides present in a spent fuel from SMR technology. The focus was mainly put on studying the physics of the system, analyzing the feasibility of reducing the inventory and radiotoxicity of the system by this method, without considering technological aspects and limitations.
\end{abstract}


\end{frontmatter}

\section{Introduction}
\label{Sec:Introduction}

In 2020, the International Atomic Energy Agency (IAEA) reported that 172 Nuclear Power Plants (NPPs) have been permanently shut down and 20 have been fully decommissioned \cite{IAEA:2023}. During the decommissioning and disposal phases, the main concern must be on the management of Spent Nuclear Fuel (SNF) due to its high and long-term radiotoxicity. 
The composition of SNF varies depending on the reactor design and the fuel burning. Thermal reactors, which operate using a slower neutron spectrum, tend to generate more minor actinides (such as Americium, Curium, and Neptunium) than reactors based on a fast neutron spectrum. Indeed, at thermal neutron energies, the likelihood of neutron absorption is higher, leading to the production of transuranic elements. As a result, the composition of SNF from these reactors typically consists of approximately 94.5\% uranium, 4.3\% fission products, 1\% plutonium (Pu), and 0.2\% minor actinides (MAs) \cite{MCFARLANE2004351}\footnote{Ratios between elements in SNF depends on the burnup level at the discharge of the reactor. Reported values are representative of SNF at 50 MWd/kg.}. Concerning the fast reactor fleet, the evaluation of typical spent fuel data is not trivial, since most of the fast technology adopts a closed fuel cycle with processed fuel, mixing MAs with fresh fuel~\cite{ORNL:2024}.\\

Currently, the vast majority of nuclear power is generated using Light Water Reactor (LWR) technology, which falls under the category of thermal reactors. 
Recent research led by Stanford University and the University of British Columbia has found that small modular reactors (SMRs), despite being promoted as a more sustainable nuclear energy option, may actually produce more radioactive waste than conventional reactors \cite{stanford:2022}. According to a study published in Proceedings of the National Academy of Sciences, SMRs could generate 2 to 30 times more waste per unit of energy produced. This poses significant challenges for waste management and disposal, as the radiotoxicity of spent nuclear fuel, particularly plutonium, from SMRs could be at least 50\% higher than that of conventional reactors after 10,000 years.
The by-product of fission (the so-called Fission Products, FP) varies in a large number of elements (hundreds of different species), the vast majority of which quickly decay into stable or long-lived elements \cite{OLANDER20014490}. Therefore, most of them do not require much effort in terms of radiological security and disposal. However, the minor part of them, the so called LLFPs (Long Lived Fission Products), together with MAs and Pu, constitute the major radiotoxicity source within the SNF, hence representing the main issue in the waste management and final disposal. Their hazardousness is due to high activity , as well as long decay time and high decay heat \cite{Office:2018}. 
To address the issue of SNF, the current established solution is geological disposal, i.e. a long-term storage inside underground engineered facilities embedded within stable geological formations. \cite{benke2023finland, onkalo2023}. \\

Given the increasing volume and complexity of waste from advanced reactors, new methods of managing nuclear waste may need to be investigated in order to evaluate their effectiveness in mitigating long-term risks. Some decades ago, Rubbia et al. proposed a road map for the treatment of spent fuel in alternative to the geological disposal \cite{Rubbia:2001}. Several techniques and designs were analysed, among which the most promising rely on the so called \textit{Partitioning and Transmutation} (P\&T) strategies \cite{OECD:2006, Kooyman:2021}. While partitioning encloses the technological processes that allow the recovery of fissile material, MAs and FPs from SNF, transmutation of long-lived fission products converts these elements into less harmful forms exploiting nuclear reactions. Nowadays, transmutation strategies are meant to be adopted in fast-fission reactors, such as some of the most promising Gen IV nuclear reactor concepts, i.e., the lead cooled fast reactor, the molten salt fast reactor, and accelerator driven systems  \cite{ALFRED:2013, MSFR:concept, Rubbia:2001, Ridikas:2002}.\\

Regarding transmutation strategies, the vast majority of designs discussed in the literature focus on \textit{neutron-based transmutation}, which rely on high neutron fluxes along with a fast neutron spectrum. This creates a scenario where the main driver for the consumption of minor actinides (i.e., neutrons) also contributes to their production. Consequently, there are no viable alternatives for SNF generated by thermal reactors. This limitation has prompted researchers to investigate non-neutron transmutation strategies, including facilities that generate high-energy particles other than neutrons. Recent advancements in photonics and ion accelerators present promising opportunities, enabling the production of energetic non-neutron particles at national lab user facilities or university accelerators. In addition, the study of alternative systems to fast
reactors for transmutation is attractive because of the lower engineering complexity that an accelerator requires compared to a nuclear plant. However, it is fair to point out that the difficulty in building a nuclear plant is repaid by the energy production, whereas in an accelerator-based scenario, there is no energy production but only consumption.
Some studies in the literature have explored non-neutron interactions for the transmutation of LLFPs. Chen et al. \cite{Chen:2008} investigated photon-based irradiation on $^{137}$Cs and $^{129}$I, while Ledingham et al. \cite{Ledingham:2010} and Wang et al. \cite{Wang:2017} applied laser-driven photons to iodine samples and $^{126}$Sn, respectively. Matsumoto et al. \cite{Matsumoto:1988} studied photon-driven beams on $^{90}$Sr and $^{137}$Cs, and finally, Wickert et al. \cite{Wickert:2024} proposed a proton-based transmutation technique. The success of these alternative methods is largely dependent on factors such as the energy and flux of the incident particles, capture cross-sections, and the design of transmutation systems. \cite{Wickert:2024} \\ \\

\begin{figure}[ht!]
    \centering
    \includegraphics[width=0.75\linewidth]{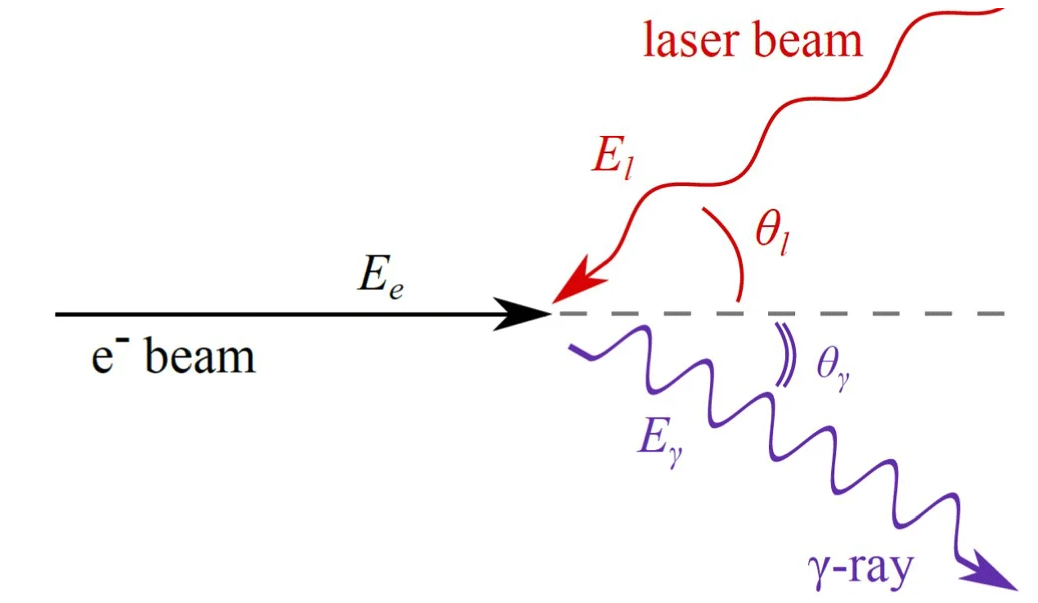}
    \caption{Schematic presentation of the photon–electron scattering process in the laboratory frame. Figure from \cite{Balabanski:2024znh}.}
    \label{fig:lcb}
\end{figure}

Currently, two primary methods exist for generating gamma rays to trigger photonuclear reactions for nuclear waste transmutation: Bremsstrahlung radiation and Compton backscattering \cite{Chen:2008}. Each method has distinct characteristics and applications, primarily differentiated by the energy distribution of the produced photons. \\
Bremsstrahlung is the simplest and most widely used method for generating high photon fluxes. This process, induced by the deceleration of intense electron beams in high-Z materials like tungsten, lead, tantalum, or depleted uranium, generates photons with a continuous energy distribution. Materials with higher proton numbers are particularly effective as converters due to their lower critical energy thresholds. Bremsstrahlung facilities are invaluable for generating intense photon beams with straightforward technology, finding applications in experiments such as the search for fission isomers and the production of neutron rich radioactive ion beams~\cite{Diamond:1999}. Despite these advantages, Bremsstrahlung sources have significant limitations, including broad energy spectra, large background signals, and lack of polarization. However, these drawbacks are negligible in programs where the focus is on beam intensity rather than energy precision \cite{Balabanski:2024znh}.\\
In contrast, recent advancements have enabled the production of high-intensity, quasi-monoenergetic gamma beams through laser Compton backscattering (LCB). This process, illustrated in Fig. \ref{fig:lcb}, involves the interaction of laser photons (energy E=h$\omega$) with GeV-level accelerated electrons: the photons gain energy from the relativistic recoil interaction, with the resulting energy scaling quadratically with the Lorentz factor of the electrons. In this manner, gamma-ray beams in the GDR energy region (10–20 MeV) can be achieved starting with laser photons of relatively low energy.  The ability to produce photon beams with a narrow energy bandwidth, strong polarization, and a high degree of tuning makes this method particularly attractive for experiments requiring precise energy control, e.g. for precise evaluation of the GDR region cross section of target nuclei. \\
Other innovative approaches include leveraging virtual photons from electromagnetic interactions involving accelerated electrons or relativistic heavy ions \cite{Balabanski:2024znh} or the usage of positron annihilation for producing quasi-monochromatic gamma rays.

Based on the above considerations, in this work a methodology to investigate the photon-driven transmutation of key MAs is developed, having focus on the fundamental physical processes involved in the MAs transmutation and proposing a new strategy for radiotoxicity reduction without relying on neutrons as the primary driver. Specifically, we explore photonuclear interactions for inducing transmutation and even fission by leveraging the Giant Dipole Resonance (GDR) in nuclei. The focus is on photo-absorption reactions, such as photocapture and photofission, triggered by high-energy photons, that have the maximum probability of occurring within the GDR region. To have a complete indication of the system's behaviour concerning the source, both photons emitted by Bremsstrahlung and LCB are considered. 
The considered MAs are a subset of the radioactive waste produced by a fuel assembly from the NuScale reactor design \cite{NuScale}, one of the most advanced and commercially ready SMR technologies. Specifically, the content of actinides present at the reactor discharge is considered separable into two groups: elements that can be \textit{recycled} for power generation purposes and those that cannot. The former, such as U and Pu, for example, are not considered to be subject to irradiation, since they can be used in so-called MOX fuels; the latter, such as Np, Am and Cm, which are the main contributors to radiotoxicity, are subject to irradiation. 
Due to the transmutations that can occur on Np, Am and Cm, this analysis considers the build-up of up to 53 nuclides. It is important to emphasize that the purpose of this work is not to present a general and industrially competitive strategy, but rather to test the feasibility on the purely physical side of minor actinide transmutations from photons, also identifying the order of magnitude of the fluxes needed to reduce the system radiotoxicity. For this reason, similar to the work of Willat et al.~\cite{WILLAT:2023} the discussion regarding the study of the costs of producing these fluxes is deliberately not considered in this paper: the initial average fluxes in the irradiated volume were taken equal to those reported by Matsumoto et al.~\cite{Matsumoto:1988}, who performed a similar work, but focusing on some LLFPs.

The present work is structured as follows: in Section~\ref{sec:methodology}, the adopted methodology is presented. Section~\ref{sec:casestudy} describes in detail the considered case study. Following, Section~\ref{sec:results} discusses the main findings, showing the evolution MAs present in a spent fuel along with the radiotoxicity. Finally, Section~\ref{sec:conclusions} resumes the main outcomes of this study, offering insights for future developments.

\section{Methodology}
\label{sec:methodology}
This section considers the basic principles for the transmutation of spent fuel if subjected to a photon beam. It is important to highlight that, within this work, with \textit{spent fuel} it is intended only the minor actinide content present at reactor discharge, leaving the contribution of fission products for future and dedicated analysis. Three elements are required: i) the \textit{nuclear data}, ii) the \textit{initial MA content} and iii) the \textit{volume averaged energy flux}.\\

First of all, nuclear data are considered, such as cross sections involved for photons, neutrons, and decay constants (see Sec.~\ref{sec:XS}) that are needed for the reaction rate evaluation. The presented methodology is totally general and can be applied to waste coming from any nuclear reactor fleet. However, the initial MAs content and the volume-averaged energy flux are by definition case dependent, so that, for sake of clarity, they will be presented along with the detailed description of the reactor design in Section~\ref{sec:casestudy}. \\
These three elements are needed for the Bateman equation solution (see Sec.~\ref{sec:bateman}), which predict the time evolution of interesting nuclide concentrations in the spent fuel knowing their initial condition as well as the irradiation flux. Once the behaviour of the spent fuel is known, the decrease in radiotoxicity given by the treatment can be evaluated (see Sec.~\ref{sec:radiotoxicity_met}). 

\subsection{Cross section data}
\label{sec:XS}
Photo-nuclear reactions are fundamental processes in which atomic nuclei interact with high-energy photons, typically high-energy X-rays, resulting in various nuclear transformations. These reactions include photo-disintegration, where a nucleus absorbs a photon and subsequently emits nucleons such as neutrons or protons, and photo-excitation, where the nucleus is excited to a higher energy state \cite{Balabanski:2024znh}. These latter two phenomena appear to have different probability of occurrence depending on the energy of the impinging photon on the nucleus. This probability, encoded in the the photonuclear absorption cross section, can be divided in three main regions\cite{Bengt:1972}, depending on the photon energy. The GDR region ($E_\gamma <$ 50 MeV), the Quasideuteron region ($50 < E_\gamma <$ 200 MeV) and the Photomeson region ($E_\gamma >$ 200 MeV). The GDR region is interesting for requiring photon energies lower than 50 MeV and exhibiting high cross sections due to resonant effects. Also, for MAs, this region begins at approximately 5 MeV. More details about the photofission mechanism can be found in Balabanski et al.\cite{Balabanski:2024znh}.\\
The knowledge of nuclear data represents the first step in this analysis. In order to exploit reactions in the GDR region, photons falling within the range 5-40 MeV are considered, along with neutrons within 0.01-20 MeV range for taking advantage of fast fission. The most significant reactions for minor actinides transmutation are summarised in the following:
\begin{itemize}
    \item Photon driven reactions: $(\gamma, f), (\gamma, n), (\gamma, 2n)$;
    \item Neutron driven reactions: $(n, f), (n, \gamma), (n, 2n)$.
\end{itemize}

\begin{figure}[htp]

    \subfloat[]{%
    \includegraphics[clip,width=\columnwidth]{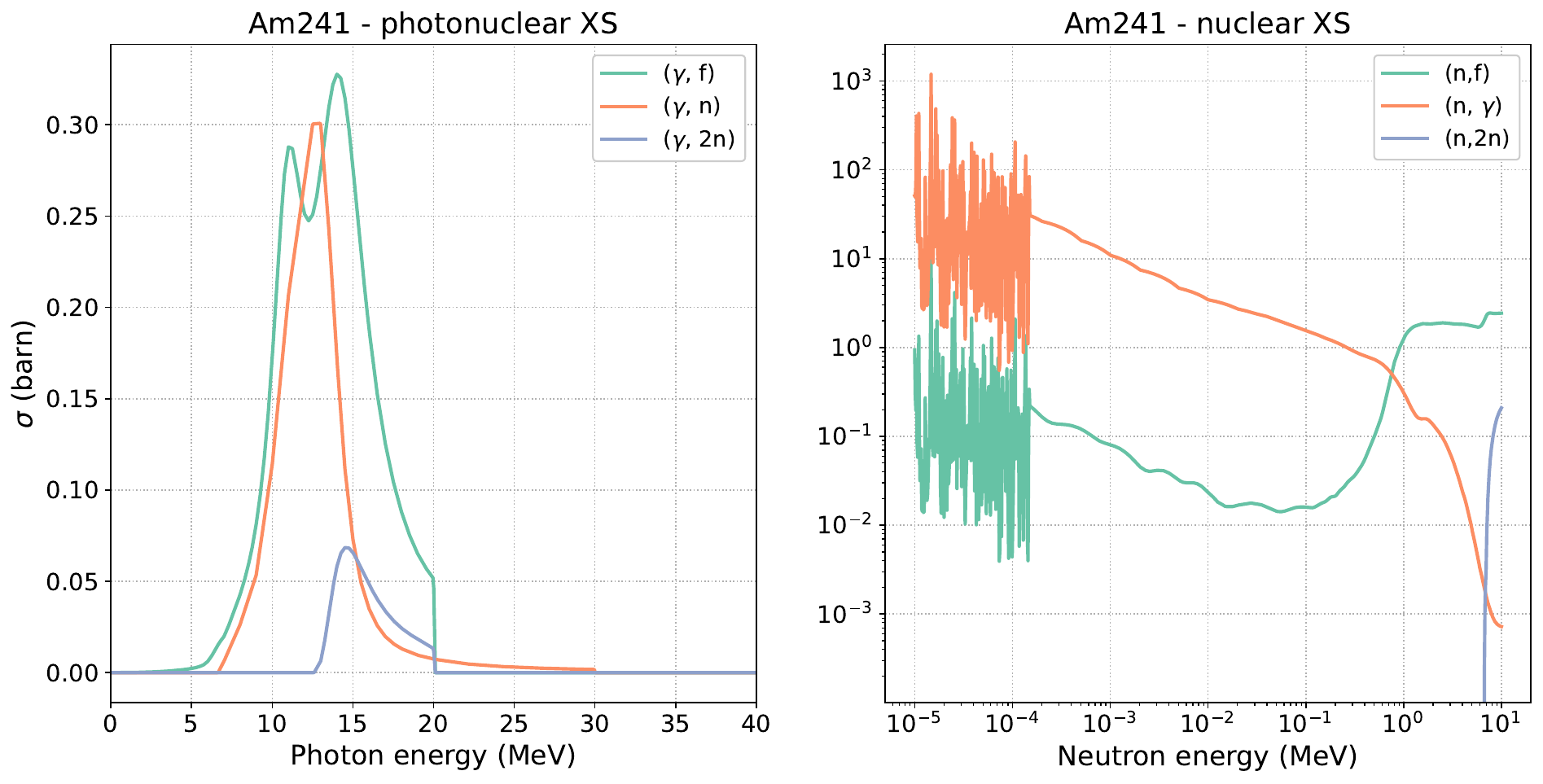}%
    \label{fig:Am241_xs}
    }

    \subfloat[]{%
    \includegraphics[clip,width=\columnwidth]{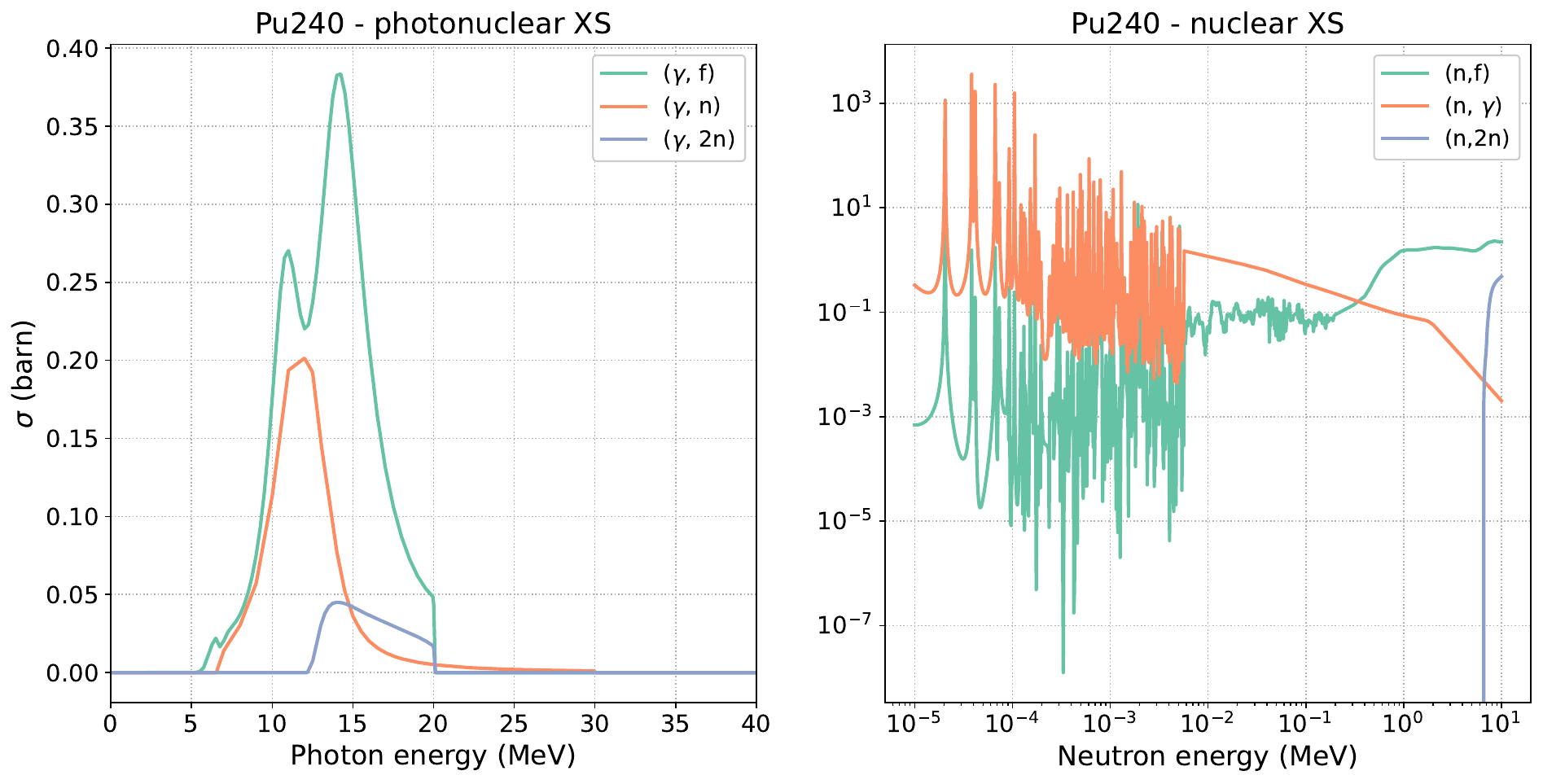}%
    \label{fig:Pu240_xs}
    }

    \caption{Photon and neutron microscopic cross section of Americium 241 (Fig.~\ref{fig:Am241_xs}) and Plutonium 240 (Fig.~\ref{fig:Pu240_xs}). When a photon impinges on the $^{241}$Am present in the spent fuel sample, this will eventually fission or being transmuted to $^{240}$Am, which rapidly become $^{240}$Pu, being successively fissioned by another photon.}
    \label{fig:Am_both_xs}

\end{figure}

When a photon with an energy in the considered range impinges on a minor actinide, it can eventually induce a photofission reaction or a transmutation like $(\gamma,n)$ or $(\gamma,2n)$, modifying the spent fuel inventory. In case of direct photofission, the actinide is burnt; while, in case of transmutation, a new nuclide is produced, being eventually disintegrated by further interactions.\\
Figure~\ref{fig:Am241_xs} illustrates the reaction cross sections for the key isotope $^{241}$Am. Consider the example where a photon with energy in the range of approximately 5 to 20 MeV interacts with a nucleus of$^{241}$Am: between 5 and 15 MeV, the photofission process competes with the $(\gamma,n)$ reaction, whereas the likelihood of the photofission reaction increases significantly above 15 MeV. Consequently, $^{241}$Am can undergo fission or be transmuted into $^{240}$Am. In the latter case, $^{240}$Am decays in $^{240}$Pu with a half-time of $\sim$ 50 h. Subsequent photon interactions are more likely to induce fission in $^{240}$Pu (see Figure~\ref{fig:Pu240_xs}), thereby reducing the inventory of spent fuel and its associated radiotoxicity.

\begin{table}[ht!]
    \centering
    \renewcommand{\arraystretch}{1.3}
    \begin{tabular}{p{1cm} p{3.25cm} p{3.25cm} p{3.25cm}}
        \hline
         & \textbf{JENDL-5} & \textbf{TENDL-2021} & \textbf{ENDF/B-VIII.0}  \\ \hline
        $(\gamma,f)$ & $^{232-237, \, 242-244}$Pu, $^{232-236, 238-244}$Np, $^{232, 237, 239-240}$U, $^{234-240, \, 242-245}$Am, $^{240-245}$Cm 
        & - 
        & $^{238-241}$Pu, $^{233-236, 238}$U,
        $^{241}$Am, 
        $^{237}$Np
         \\ \hline 
        $(\gamma,n)$ & - 
        & All 53 nuclides
        & - \\ \hline
        $(\gamma,2n)$ & - 
        &  $^{232,237, 239-240}$U, $^{232-236,238-244}$Np, $^{232-237,242-244}$Pu, $^{234-240,242-245}$Am, $^{240-245}$Cm 
        & $^{238-241}$Pu, $^{233-236, 238}$U, $^{237}$Np,  $^{241}$Am  \\ \hline
        $(n,f)$ & - 
       &   $^{232-233, 240-244}$Np, $^{232-235}$Pu, $^{234-239,245}$Am
        & $^{232-240}$U, $^{234-239}$Np, $^{236-244}$Pu, $^{240-244}$Am, $^{240-245}$Cm  \\ \hline
        $(n,\gamma)$ & - 
        & $^{232-233, 240-244}$Np, $^{232-235}$Pu, $^{234-239,245}$Am
        & $^{232-240}$U, $^{234-239}$Np, $^{236-244}$Pu, $^{240-244}$Am, $^{240-245}$Cm \\ \hline
        $(n,2n)$ & - 
        & $^{232-233, 240-244}$Np, $^{232-235}$Pu, $^{234-239, 245}$Am
        & $^{232-240}$U, $^{234-239}$Np, $^{236-244}$Pu, $^{240-244}$Am, $^{240-245}$Cm \\ \hline
    \end{tabular}
    \caption{Classification of nuclear data for employed reactions}
    \label{tab:XS_library}
\end{table}

Finally, this work relies on a diverse array of nuclear data sources, given the lack of a single comprehensive database containing all relevant reaction information. The considered libraries are JENDL-5 \cite{JENDL5}, TENDL-2021 \cite{TENDL} and ENDF/B-VIII.0 \cite{ENDF8}. Table~\ref{tab:XS_library} classifies the adopted nuclear data libraries for each considered reactions for the 53 isotopes. It is evident that the data were preferably taken from ENDF/B-VIII.0 and, when necessary, were completed with data taken from the other two libraries. A complete analysis should also consider the uncertainties related to each cross section and propagate the error in successive calculations. However, at the time this analisys was performed, the only nuclear library that  contains covariance matrices is TENDL-2023. We plan to perform verification analysis relying on that nuclear library only, to also get the uncertainty propagation on the results.

\subsection{Bateman equation solution}
\label{sec:bateman}

The information of the cross section alone is not sufficient to evaluate the transmutation capability of photons against neutrons. To do so, it is also crucial to compare the absolute values of reactions such as $(\gamma,f)$ and $(n,f)$ for neutrons in the energy range above 0.5 MeV (see Figure~\ref{fig:Am_both_xs}). In this energy range, when neutron and photon fluxes are of equal intensity, the probability of neutron interactions is approximately three times higher than that of photon interactions. However, the actual intensities of these two fluxes differ significantly. The neutron flux is generated as secondary particles during photonuclear interactions and fission, while the photon flux originates from an accelerator, which produces a known and controllable beam. A more complete discussion about fluxes will be faced in Section~\ref{sec:casestudy}, showing how to extract volume averaged energy fluxes (i.e., $\Phi_\gamma(E), \Phi_n(E)$) from the system and how the normalization was set in this work.

\begin{figure*}
    \centering
    \includegraphics[trim={0 0 0 0},scale=0.504, angle=90]{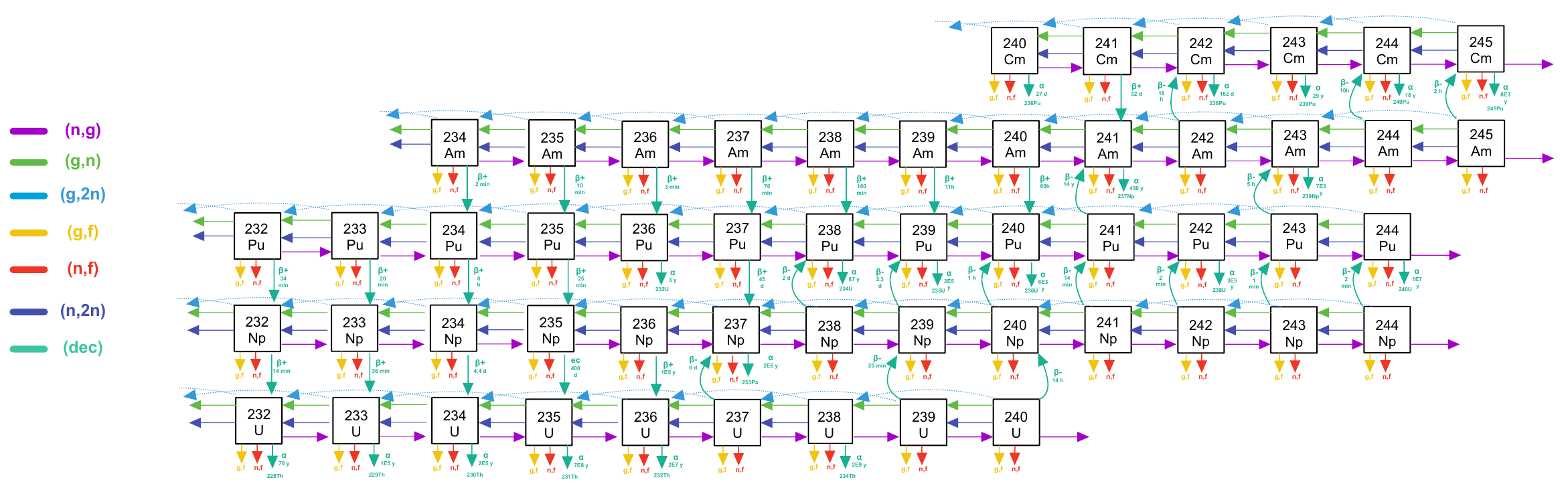}
    \caption{Scheme for the interaction between the considered 53 nuclides, showing the possible chain reactions from neutrons, photons and decay.}
    \label{fig:TransmutationMap}
\end{figure*}

The Bateman equations describe the time evolution of a set of nuclides undergoing radioactive decay and transmutation due to various interactions. In particular, this work considers 53 isotopes in the minor actinides reaction chain. When considering minor actinides subjected to a photon flux, it is essential to account for both neutron as well as photon interactions: indeed, photonuclear interactions comes directly from the primary beam, while neutron interactions are possible thanks to their generation in fission and $\left(\gamma,n \right)$ channels.
Fig~\ref{fig:TransmutationMap} is a graphical representation of the Bateman equations, reporting the map of the considered nuclei, highlighting their links due to photon/neutron reactions and radioactive decay. 

The system is solved using a lumped approach, which means that no spatial discretization is applied. Instead, the driven fluxes are considered as volume-averaged quantities. This method provides a comprehensive overview of the mass evolution.\\
Equation~\eqref{eqn:bateman} shows the analytical formulation of the Bateman, represented by a balance between creation/destruction of the isotope $i$ among its own channels and the reactions involving other isotopes $j$:

\begin{equation}
\label{eqn:bateman}
\begin{split}
    \frac{dN_i(t)}{dt} &= \underbrace{-\lambda_i N_i(t)}_{\text{destruction by decay}} 
    + \sum_{j \neq i} \underbrace{\lambda_{j\to i} N_j(t)}_{\text{creation by decay}} \\
    & \underbrace{- N_i(t) \int \left[ \sigma_{\gamma,f}^i(E) + \sigma_{\gamma,n}^i(E) + \sigma_{\gamma,2n}^i(E)\right] \Phi_\gamma(E) \, dE}_{\text{destruction by photon capture}}  \\
    & + \sum_{j \neq i} \underbrace{ N_j(t) \int \left[ \sigma_{\gamma,n}^{,j \to i}(E) + \sigma_{\gamma,2n}^{,j \to i}(E) \right]\Phi_\gamma(E) \, dE }_{\text{creation by photon reactions}} \\
    & \underbrace{- N_i(t) \int \left[ \sigma_{n,f}^i(E) + \sigma_{n,\gamma}^i(E) + \sigma_{n,2n}^i(E)) \right] \Phi_n(E) \, dE}_{\text{destruction by neutron capture}}  \\
    & + \sum_{j \neq i} \underbrace{ N_j(t) \int \left[ \sigma_{n,\gamma}^{,j \to i}(E) + \sigma_{n,2n}^{,j \to i}(E) \right]\Phi_n(E) \, dE }_{\text{creation by neutron reactions}} \\
\end{split}
\end{equation}

Where $N_i$ represents the atomic density of nuclide $i$ in unit of atom/cm$^3$. 

Equation~\eqref{eqn:bateman} appears to be linear and easily solved through the exponential matrix inversion only if the energy fluxes $\Phi_\gamma$ and $\Phi_n$ are not time dependent. This hypothesis holds only if the change in atomic density is small relative to the initial condition; namely, a large change in atomic density induces a perturbation volume averaged fluxes, due to differences in attenuation. \\For this reason, this work adopts a forward Euler scheme to take into account the modifications in the particle fluxes during time. This method is simple but effective: first, the time frame in which the equation will be solved is divided in substeps. During the first substep, the system of equations is solved with the spent fuel initial condition and with the initials volume averaged fluxes kept constant. At the end of the substep, the final composition is recorded and it is adopted for the re-evaluation of the fluxes. Then,  the solution can proceed with the second substep, up to the end of the time frame. The presented algorithm is reported for clarity in Algorithm~\ref{algo:euler}.

\begin{algorithm}
\caption{Forward Euler Method for Solving $\frac{dN}{dt} = B(t, N)$}
\begin{algorithmic}[1]
\label{algo:euler}
\STATE \textbf{Input:} Initial time $t_0$, initial value $N_0$, step size $h$, number of steps $H$
\STATE \textbf{Initialize:} $t \gets t_0$, $N \gets N_0$
\FOR{$k = 0$ to $H - 1$}
    \STATE Compute derivative: $B_n \gets B(t, N)$
    \STATE Update value: $N \gets N + h \cdot B_n$
    \STATE Advance time: $t \gets t + h$
\ENDFOR
\STATE \textbf{Output:} Approximate solution $N$ at $t = t_0 + H \cdot h$
\end{algorithmic}
\end{algorithm}

\subsection{Radiotoxicity evaluation}
\label{sec:radiotoxicity_met}
Contents of radioactive isotopes in spent fuel can be assessed in terms of mass, activity, energy release or radiotoxicity. The latter figure of merit gives quantitative information on the radiobiological influence on the human body. Hazard from minor actinides is mainly due to alpha decay, which having long half-time, require strategies for their disposal or treatment.\\
The radiotoxicity of a single nuclide is defined as the product between the \textit{activity} (i.e., the number of disintegration per second, measure in Becquerel) and the \textit{effective dose coefficient} (i.e., a measure of the damage done by ionising radiation on an organism, measured in Sievert per Becquerel)\footnote{The Sievert (Sv) is a measure of the dose arising from the ionisation energy absorbed, describing the biological effect of radiation deposited in an organism. The biological effect of radiation is not just directly proportional to the energy absorbed in the organism but also by a factor describing the quality of the radiation \cite{Rubbia:2001}}. Among various definitions of effective dose coefficient, the most popular describe the damage done by \textit{ingestion} or by \textit{inhalation}. This work chose to focus on the former, keeping in mind that the same methodology would apply also in the inhalation scenario. \\\
Also, we focused to the radioactivity content coming from MA only, without considering production/transmutation of LLFP that may build up during irradiation.
The lumped approach used in this work ensures the estimation of isotopic concentration evolution, allowing for the calculation of radiotoxicity. In particular, for a system of $M$ nuclides, described by their own decay constant $\lambda_i$, radiotoxicity of the system is calculated as:
\begin{equation}
\label{eqn:radiotoxicity_system}
    R_{sys}(t)= V_p \sum_{i=1}^M y_i \lambda_i N_i(t) 
\end{equation}
where $V_p$ is the irradiated volume where actinides are uniformly distributed, in cm$^3$, $y_i$ are the dose coefficients, tabulated within the ICRP 2012 report \cite{ECKERMAN20121} and $N_i$ is the outcome concentration of the Bateman equations in at/cm$^3$. 

The system radiotoxicity $R_{sys}$ is compared with the \textit{free evolution radiotoxicity} $R_{free}$, namely the Sievert emitted if the same mass of minor actinides used as the initial condition for this calculation (i.e., $N_i(0)$) were to evolve without any particle flux acting as a driver. This concentration evolution has been evaluated with an independent Serpent simulation adopting the \textit{decay step mode}, where Bateman equations are internally solved from the code, therefore obtaining the concentrations $N_i^{free}$:

\begin{equation}
\label{eqn:radiotoxicity_free}
    R_{free}(t)= V_p \sum_{i=1}^M y_i \lambda_i N_i^{free}(t) 
\end{equation}
It is worth noticing that decay constants adopted for the evaluation of $N_{i}^{free}$ were taken from ENDF/B-VIII.0\cite{endfviii}.
Both the quantities $R_{sys}$ and $R_{free}$ have been normalised with respect to the natural uranium radiotoxicity $R_{nat}$, calculated as 
\begin{equation}
\label{eqn:radiotoxicity_natural}
    R_{nat}= V_{core} \left(N_{U238}^{nat} \lambda_{U238} y_{U238} + N_{U235}^{nat} \lambda_{U235} y_{U235} + N_{U234}^{nat} \lambda_{U234} y_{U234} \right)
\end{equation}

Where $V_{core}$ represents the volume associated with the initial reactor core fuel material, and the concentrations considered are those of natural uranium. This normalization accounts for the time it takes for the radiotoxicity to decrease compared to the value that the same amount of fuel would have if it were considered natural uranium.

\section{Case study}
\label{sec:casestudy}

The spent fuel inventory is primarily influenced by both the reactor design and the nominal power. The reactor design affects the composition of fresh fuel, including the types of enrichment and fissile materials present, as well as the evolution of the neutron energy spectrum. For instance, a reactor operating with a thermal spectrum tends to burn odd nuclei more quickly, while a reactor with a fast spectrum can effectively burn both odd and even nuclei. The nominal power, on the other hand, determines the rate at which the fuel is consumed, providing insight into the number of minor actinides generated through neutron captures over the reactor lifetime. The figure of merit that encapsulates fuel history is \textit{burnup} (in MWd/kgU), which is proportional to the total energy produced by the reactor. A complete report with spent fuel inventory from different reactor technology and disposal choices could be found at~\cite{Wigeland:2015}. \\

\begin{figure}
    \centering
    \includegraphics[width=.5\linewidth]{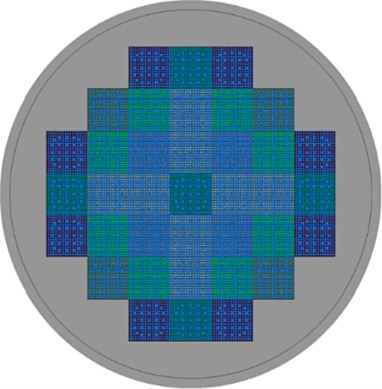}
    \caption{Schematic view of NuScale core from \cite{NuScale}.}
    \label{fig:core}
\end{figure}

The methodology presented in this paper is general, and its application is not restricted to the presented case only. However, we believe that it could be interesting for a general reader to learn the details of one of the most promising designs for the near future of energy generation from nuclear sources.
The present investigation focuses on the NuScale Power Module™ (NPM)\cite{NuScale}, which is a small, light-water-cooled Pressurized-Water Reactor (PWR) designed by NuScale Power LLC. Being one of the most mature SMR designs, it is recognized to be the first one certified for the use in the United States by U.S. Nuclear Regulatory Commission (NRC), considered to be an Early-deployable SMR technology with a nominal power of 50 MWe. The plant is modular, being able to accommodate a varying number of NPMs to meet customer’s energy demands.\\
The design consists in 37 fuel assemblies including 16 control rod assemblies that make up the core arrangement, as depicted in Figure~\ref{fig:core}. The typical 17 x 17 PWR fuel assembly with 24 guide tube sites for control rods and a central instrument tube serve as the model for the fuel assembly design. The assembly is held by five spacer grids and is supposedly half the height of a typical  plant ($\sim$2 m). For some rod locations, the fuel is UO$_2$ with Gd$_2$O$_3$ homogeneously mixed in as a burnable absorber (BA). The $^{235}$U enrichment is less than the 4.95\% maximum currently set by US manufacturers.\\
Despite its smaller size, the waste produced by the NuScale reactor shares similar challenges with other SMR designs. The compact core, operating with higher neutron flux and burnup rates, results in the generation of highly radioactive spent fuel with a complex isotopic composition.\\

\subsection{Initial condition: spent fuel composition}
\label{sec:Serpent}
In the present work, the NuScale core has been modelled and simulated with the Serpent Monte Carlo code\cite{SERPENT}. The software is suited to solve neutron transport in fissile systems, being able to simulate depletion evolution through burnup schemes: for the presented scenario, it has been hypothesized a reactor cycle from fresh fuel up to 50 MWd/kg, with 100 active cycles, 50 inactive and $10^6$ particles simulated per cycle. 
The SMR design has been reproduced with high fidelity in a 3D domain. The adopted burnup scheme was the Stochastic Implicit Euler, well suited for 3D burnup scenarios, able to stabilize the well known problem of nuclide concentration oscillations\cite{MonteCarlo_SIE_stability}. Nuclear data for neutron interactions, fission yield and decay data are taken from ENDF/B-VIII.0\cite{ENDF8}. 

To obtain the most realistic estimate of the initial condition of the spent fuel, a cooling period of two years is considered. The fuel mass inventory at 50 MWd/kg is let free to decay for such period. This cooling window allows the nuclei having the shortest half-life to undergo radioactive decay, which reduces the number of different nuclides in the system, thereby altering the inventory compared to that of freshly discharged fuel. \\
The initial conditions is based on the minor actinides present in the whole NuScale fuel, shown in Table~\ref{Table:InitialComposition}. It is important to note that uranium and plutonium were deliberately excluded from this calculation, as their reuse in MOX fuels is widely expected in the literature\cite{Bairiot:1992}. The actinides considered in this work suffers from being not viable as nuclear fuel at present due to their poor neutronic performances, moreover they are responsible for a significant radiotoxicity level (i.e., $^{241}$Am and $^{237}$Np).

\begin{table}[ht!]
\renewcommand{\arraystretch}{1.5}
    \centering
    \begin{tabular}{c|ccccccc}
        \hline
        \textbf{Isotope} &  \textbf{$^{241}$Am}    & \textbf{$^{243}$Am} & \textbf{$^{242}$Cm} & \textbf{$^{243}$Cm} & \textbf{$^{244}$Cm} & \textbf{$^{245}$Cm} & \textbf{$^{237}$Np} \\ \hline 
         
        \textbf{Mass (g)} & 2616   & 3710 &  14.08 &  10.0 & 2254 & 217.7 & 5425 \\       \hline

    \end{tabular}
    \caption{More relevant minor actinide (Am + Cm + Np) composition of spent fuel in NuScale rector assembly at 50 MWd/kg after 2 years in a pool.}
    
    \label{Table:InitialComposition}
\end{table}

\subsection{Flux evaluation in different scenarios}
\label{sec:fluka}
The interaction between a particle beam and the spent fuel is not trivial due to the high number of reaction channels available. Since this work uses a lumped approach, each isotope transmutation is driven by an average reaction rate, depending both by the microscopic cross section as well as by the average flux value. To have a precise estimation of the volume-averaged particle flux in a spent fuel sample irradiated by a particle beam, a set of numerical simulations are performed using the FLUKA  code (version 2024.1.0) \cite{BATTISTONI201510}. FLUKA is a Monte Carlo radiation transport code that offers high-precision modeling of particle interactions and transport in complex 3D geometries. It supports a wide range of particles (neutrons, photons, electrons, hadrons, heavy ions) across a broad energy spectrum, using continuous-energy cross sections and detailed physics models. FLUKA naturally handles coupled transport, energy deposition, and secondary particle production, making it ideal for radiation shielding, dosimetry, and activation analyses. Its flexible source definition, advanced variance reduction techniques, and extensive built-in scoring capabilities allow accurate and efficient solutions to complex radiation transfer problems in both stationary and time-dependent scenarios.\\
The irradiated sample is a cylinder having radius equal to 1 cm and height equal to 4 cm. This is chosen considering the attenuation of the primary beam and the secondary particles produced inside the spent fuel: the $\sim$~12 cm$^3$ cylinder delimits the zone where both the primary beam intensity and the secondary particle flux are reduced by a factor 1000. The volume is filled with a fictitious ceramic material MO$_2$, where M stands for the isotopic mix shown in Tab.~\ref{Table:InitialComposition}. The density has been set equal to 10.2 g/cm$^3$, in the same order of magnitude of common oxides adopted in the nuclear field.\\
\begin{figure}
    \centering
    \includegraphics[width=.8\linewidth]{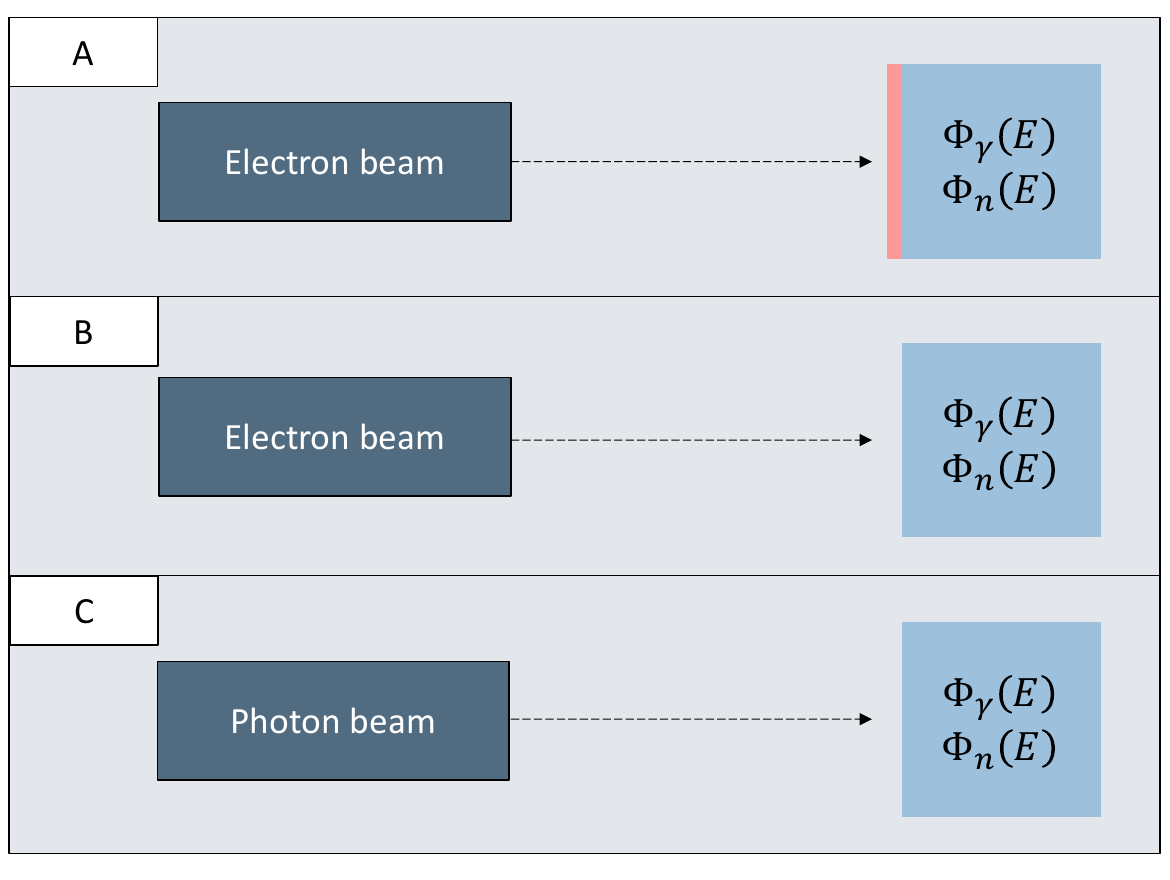}
    \caption{Scheme for the three scenarios considered in FLUKA. Photons are generated by Bremsstrahlung in cases (A) and (B), with the difference that the former adopts a Tantalum converted (red)  while the second convert electron into photons directly in the spent fuel sample. Case (C) assumes a photon beam from LCB. In all the three cases, the FLUKA calculation evaluates the volume-averaged photon and neutron energy flux.}
    \label{fig:FluxScheme}
\end{figure}
The proposed methodology has been tested with two source configurations, namely photons generated by Bremsstrahlung or by LCB. The considered scenarios are graphically represented in Fig.~\ref{fig:FluxScheme}. 
An input beam (dark blue box) impinges on the spent fuel sample (light blue box), then:
\begin{itemize}
    \item \textbf{Simulation A} considers electrons with 40 MeV energy on a Tantalum converter (red rectangle) for being converted into photons through Bremsstrahlung. To find the Tantalum width that maximizes the photon production, a sensitivity analysis related to the converter size is proposed, repeating the calculation for 0.5 mm, 1 mm and 2 mm Tantalum width;
    \item \textbf{Simulation B} considers the same 40 MeV electron beam, but removes the Tantalum layer, allowing the conversion to occur directly in the spent fuel material;
    \item \textbf{Simulation C} considers a Gaussian-shaped photon spectra impinging on the spent fuel sample, simulating the LCB scenario. 
\end{itemize}
The interactions between the beam and the spent fuel target enable the estimation of a volume-averaged flux responsible for nuclei transmutation. While it is recognized that photons (in photo-nuclear reactions), neutrons (in neutron interactions), and electrons (in electrofission) can induce transmutation, the latter has not been considered in this analysis due to its relatively low impact compared to the other channels \cite{Oleinikov:2023}. In all the cases presented, the energy fluxes of photons, $\Phi_\gamma (E)$, and neutrons, $\Phi_n (E)$, are computed using FLUKA with 10$^7$ primary particles. The spatial domain considered is the spent fuel, while the energy domain for photons ranges from 5 MeV to 40 MeV and for neutrons, it spans from 0.01 MeV to 20 MeV.\\

The initial normalization of such fluxes is a crucial point in the presented analysis. The fluxes calculated by FLUKA are known despite a normalization factor F\footnote{For example, the FLUKA simulation evaluates the photon flux in the unit of $\frac{\gamma}{cm^2 \cdot GeV \cdot source}$}, which depends by the source intensity of the accelerator. We considered different photon sources (i.e., bremsstrahlung and LCB) and our purpose is to compare the potentials of these sources and their effect on radiotoxicity by considering them \textit{at the same averaged flux volume}. For this reason, the normalization factor was chosen by constraining the number of $\mathrm{\frac{\gamma}{cm^2 s}}$ within the spent fuel sample. Namely, for each simulation:

\begin{equation}
    \label{eqn:flux_norm}
    \text{Find F$_s$ such that} \int \text{F}_s \cdot \Phi_\gamma(E,t=t_0)\, dE\, = \,P \,\, \left(\frac{\gamma}{\text{cm}^2 \text{s}} \right)
\end{equation}

\begin{figure}
    \centering
    \includegraphics[width=1\linewidth]{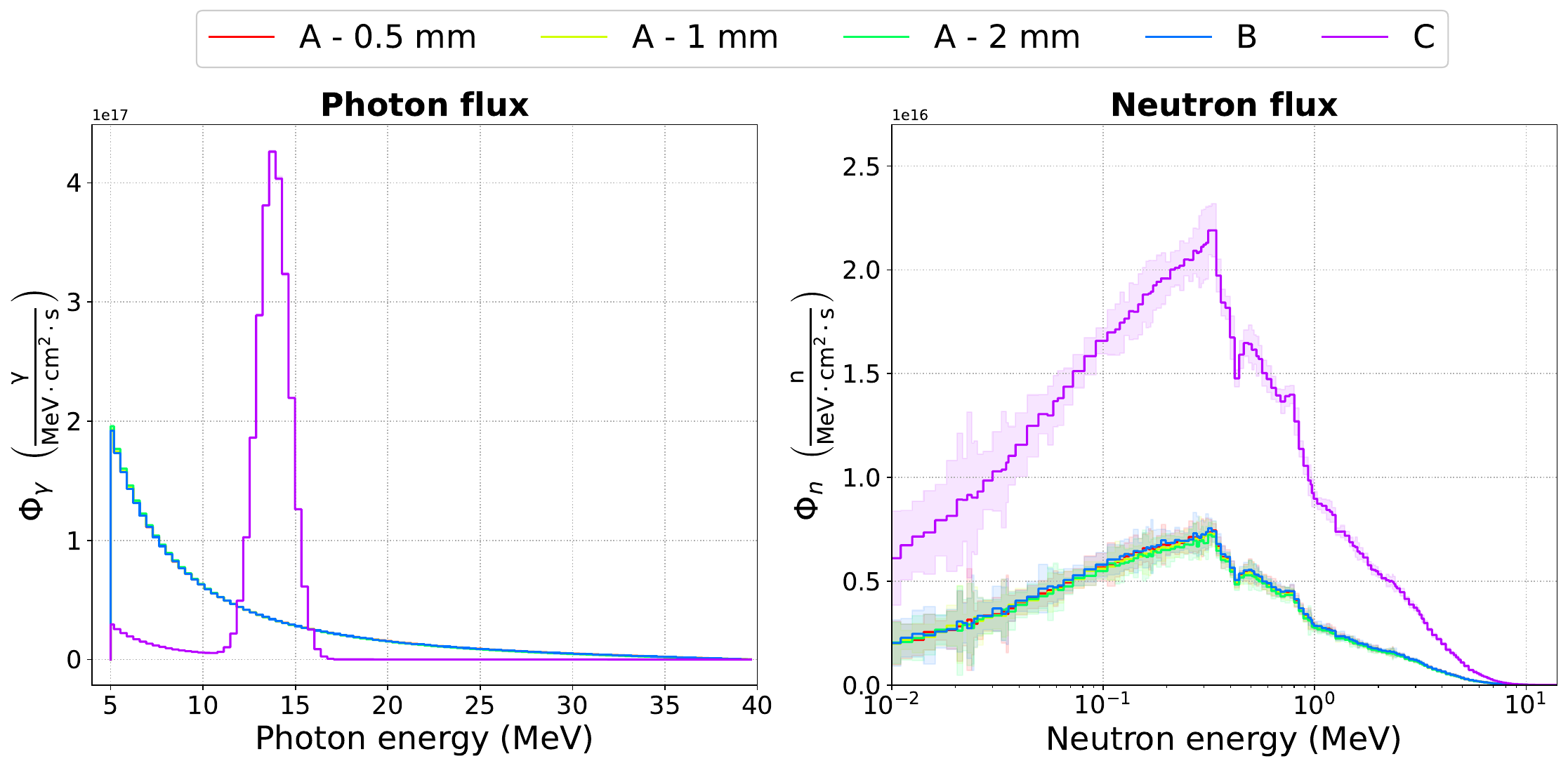}
    \caption{Photon and neutron fluxes dependent on source type. Coloured areas represent the statistical uncertainty. Normalization set to 1E+18 $\mathrm{\frac{\gamma}{cm^2 s}}$.}
    \label{fig:phiE_source}
\end{figure}

Where $P$ is the \textbf{volume-averaged flux} and $t_0$ represents the beginning of the irradiation period. The value of F$_s$ can also be adopted to normalize the neutron flux $\Phi_n(E)$.
This work considers four different flux scenarios values, i.e. 1E+17, 1E+18, 1E+19, 1E+20 $\mathrm{\frac{\gamma}{cm^2 s}}$. These values of flux are taken for reference from a similar work made by Matsumoto et al.\cite{Matsumoto:1988}. It is important to highlight that the considered values were considered for having a comparison with the work of Matsumoto et al. and may not reflect realistic source intensities, which is not the purpose of this work. In particular, different ranges of fluxes are present nowadays for LCB and Bremmstrahlung facilities, the first ranging from 1E5 $\gamma/s$ up to 1E11 $\gamma/s$, while the latter having maximum to 2E18 $\gamma/s$\cite{Balabanski:2024znh}.

Figure~\ref{fig:phiE_source} reports the scored photon and neutron fluxes with the initial spent fuel composition in the different simulations: with the adoption of the USRTRACK card, the number of tracks of photons and neutrons are recorded in an equi-spaced energy mesh. Then, the shapes are normalized according to Eqn.~\eqref{eqn:flux_norm}. It is interesting to note few differences in fluxes arising from Bremsstrahlung for diverse Tantalum thicknesses (cases A) and for no Tantalum converter (case B), whereas an LCB source (case C) is able to reproduce quasi-monoenergetic photons in the target volume and a more intense neutron field within the spent fuel.\\

The working path is now resumed here, explicating the key passages and leaving the results for the next section:
\begin{enumerate}
    \item The initial condition is set equal to the amount of Americium, Curium and Neptunium present in the SMR NuScale design at a depletion time of 50 MWd/kg;
    \item After a decay time of 2 years in a decay pool, part of the amount of MAs is extracted, manufactured in a pellet shape and irradiated by a driver flux;
    \item Three irradiation scenarios have been considered accordingly to Figure~\ref{fig:FluxScheme}, scoring in each case the photon and the neutron energy fluxes within the pellet (see Figure~\ref{fig:phiE_source});
    \item The initial value problem (IVP) reported in Eqn.~\eqref{eqn:bateman} representing the nuclide evolution subjected to photon, neutron and decay interactions is solved with the mentioned initial condition and the different driver fluxes for a time window of 100 years. Different normalization levels have been considered, setting the averaged photon flux to 1E+17, 1E+18, 1E+19 and 1E+20 $\mathrm{\frac{\gamma}{cm^2 s}}$. The system is solved with the SciPy\cite{2020SciPy} package \textit{solve\_ivp}, with the Radau method\footnote{This method is appropriate for the so-called stiff problems: since the characteristic time of each equation may be different (up to several order of magnitude), a robust strategy in solving the differential equations is required.}.
\end{enumerate} 

Finally, it's important to highlight that the considered irradiation sample is a pellet of 12 cm$^3$, whereas the total amount of minor actinide volume to treat from the NuScale facility is $\sim$1400 cm$^3$. So that, all the analysis present within this section will be about the treatment of $\sim$0.9 \% of the total waste produced in the system (i.e., $\sim 130$ g).

\section{Results and Discussion}
\label{sec:results}

\subsection{Atomic concentrations evolution}
\label{sec:atomic_concentration_results}
\begin{figure}
    \centering
    \includegraphics[width=1\linewidth]{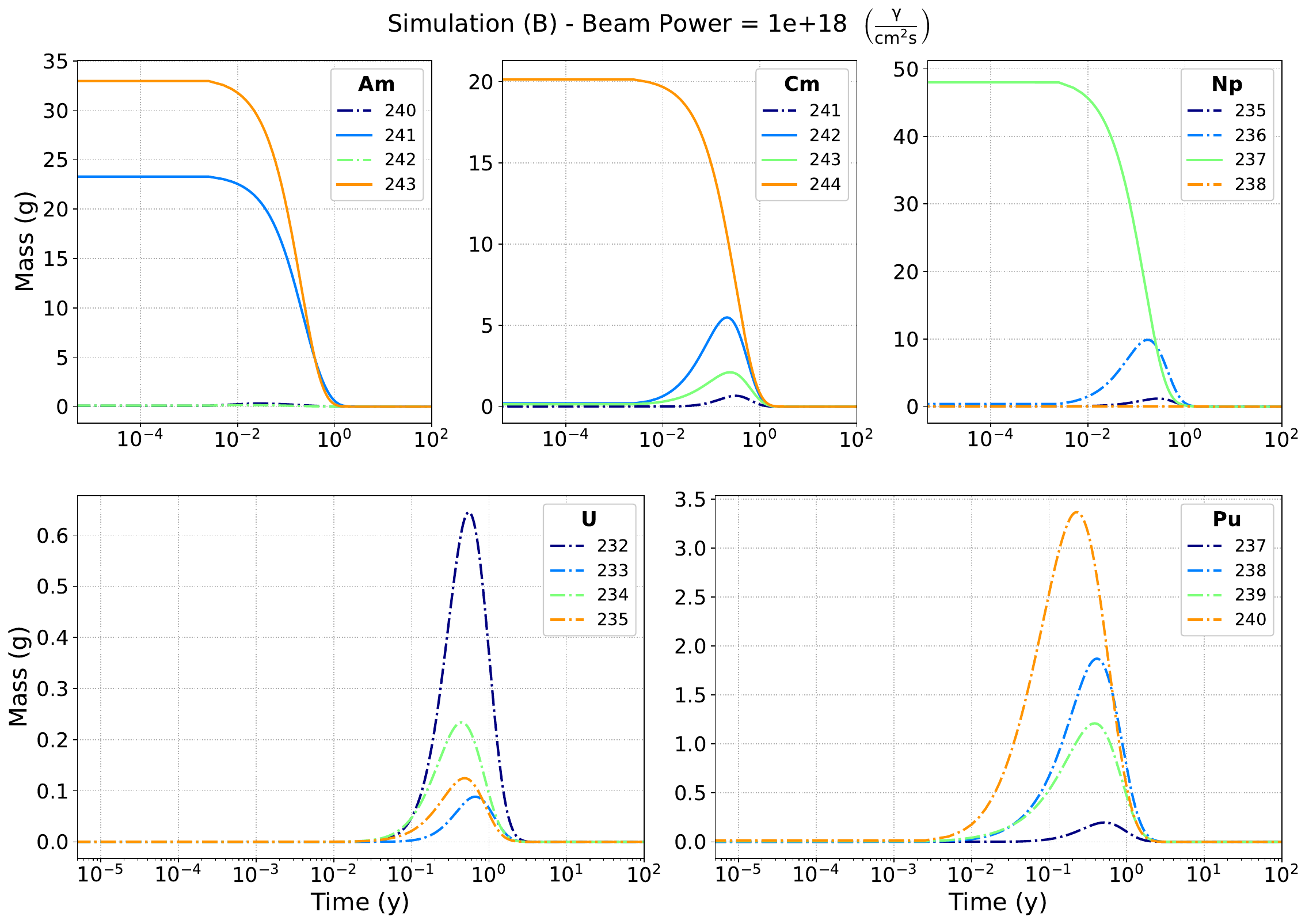}
    \caption{Minor actinide concentration in Simulation (B), considering a volume average flux of 1E+18 $\mathrm{\frac{\gamma}{cm^2 s}}$. Solid lines represent the nuclides considered in the initial condition, whereas dotted lines are the by-product.}
    \label{fig:concentrationsB}
\end{figure}

\begin{figure}
    \centering
    \includegraphics[width=1\linewidth]{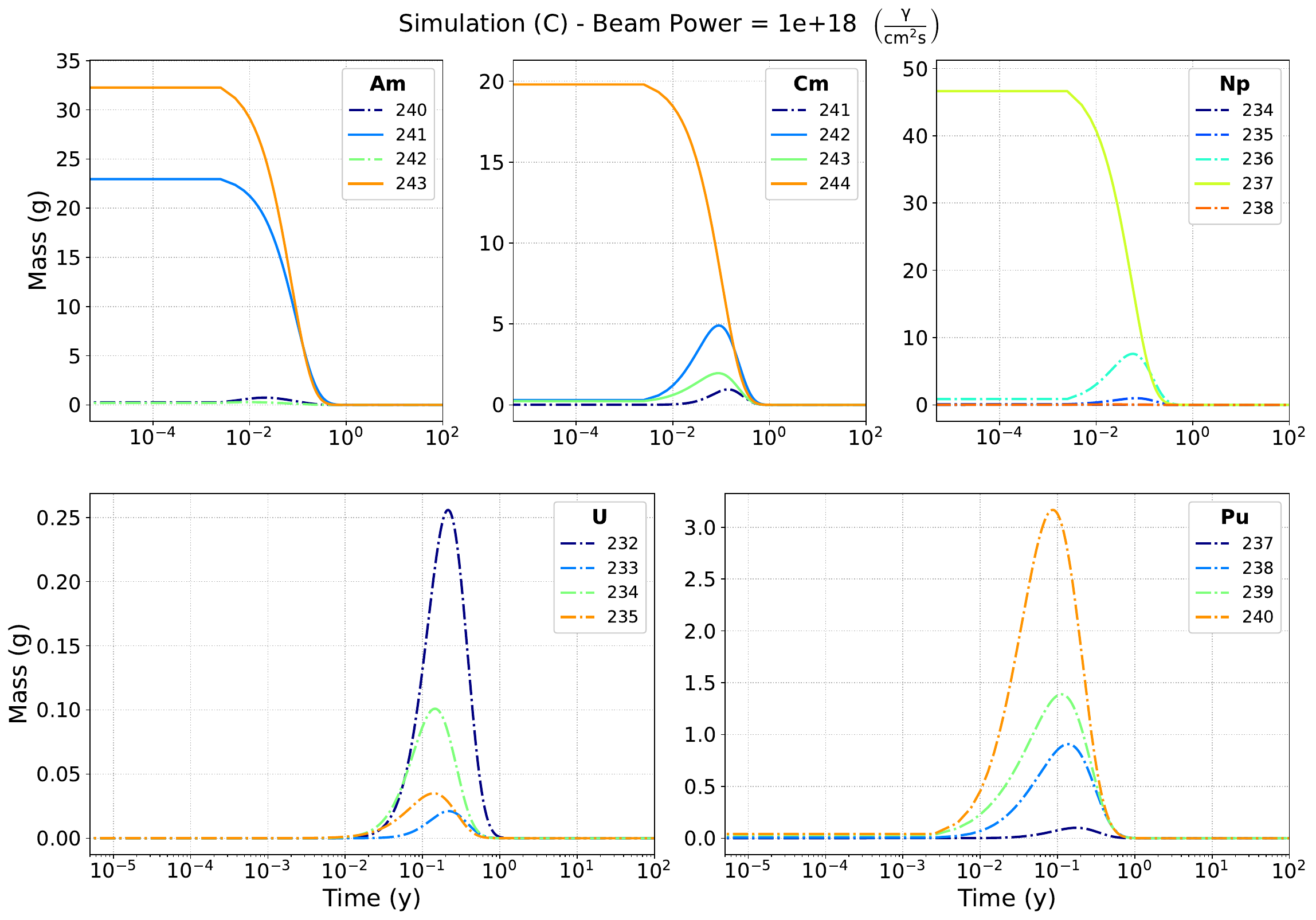}
    \caption{Minor actinide concentration in Simulation (C), considering a volume average flux of 1E+18 $\mathrm{\frac{\gamma}{cm^2 s}}$. Solid lines represent the nuclides considered in the initial condition, whereas dotted lines are the by-product.}
    \label{fig:concentrationsC}
\end{figure}

The time evolution of the concentrations are obtained for the observed states in the different considered scenarios.
From the analysis reported in Sec.~\ref{sec:fluka}, both the photon and neutron fluxes from scenarios A and B shown in Figure~\ref{fig:phiE_source} do not have appreciable differences, meaning that the photon/neutron buildup within the pellet is not sensible to a an external converter. For this reason, from now on just the results concerning scenarios B and C will be shown. The driver fluxes are updated over time with the Forward Euler scheme reported in Algorithm~\ref{algo:euler}: within the considered time frame of 100 years, we performed 5 substeps at [0.0, 0.25], [0.25, 0.50], [0.50, 1], [1, 5] and [5, 100] years where the fluxes are re-evaluated based on the change in composition. The overall nuclides' evolution in concentration is shown in Figure~\ref{fig:concentrationsB} and Figure~\ref{fig:concentrationsC} for cases B and C, respectively. Two major behaviours can be appreciated. The first involves those nuclides present in a considerable amount in the initial condition, namely $^{241}$Am, $^{243}$Am, $^{244}$Cm and $^{237}$Np (solid lines). They tend to decrease in time, being both transmuted with $(\gamma, xn)$ or $(n, \gamma)$ reactions or destroyed by fission. Then, the second behaviour involves the nuclides formed through the transmutation of the original isotopes. Several byproducts (indicated by dotted lines) appear in the system, initially starting at low concentrations, peaking when the transmutation effect of the precursors is at its maximum, and then reducing due to the combined effects of transmutation and fission. Among the 53 nuclides studied, only the four primary contributors in terms of mass for each species are presented, as the remaining nuclides are present only in traces.\\
The shown results consider an initial photon flux normalized to 1E+18 $\mathrm{\frac{\gamma}{cm^2 s}}$. Since the flux is a linear parameter in the Bateman equation, the dynamics of nuclide evolution (specifically, how they tend to evolve relative to each others) appears to be unaffected by the power level. Indeed, the consumption of the nuclides is directly proportional to the averaged flux value. For this level, the attenuation of the actinides is achieved in less than one year. A more detailed discussion regarding the quantification of this treatment's effects will be provided in the next subsection. It's interesting to notice that similar results in terms of concentration reduction were found by \cite{Matsumoto:1988} with photon flux targeting fission products.

\subsection{Radiotoxicity estimation}
\label{sec:radiotoxicity_res}

\begin{figure}
    \centering
    \includegraphics[width=1\linewidth]{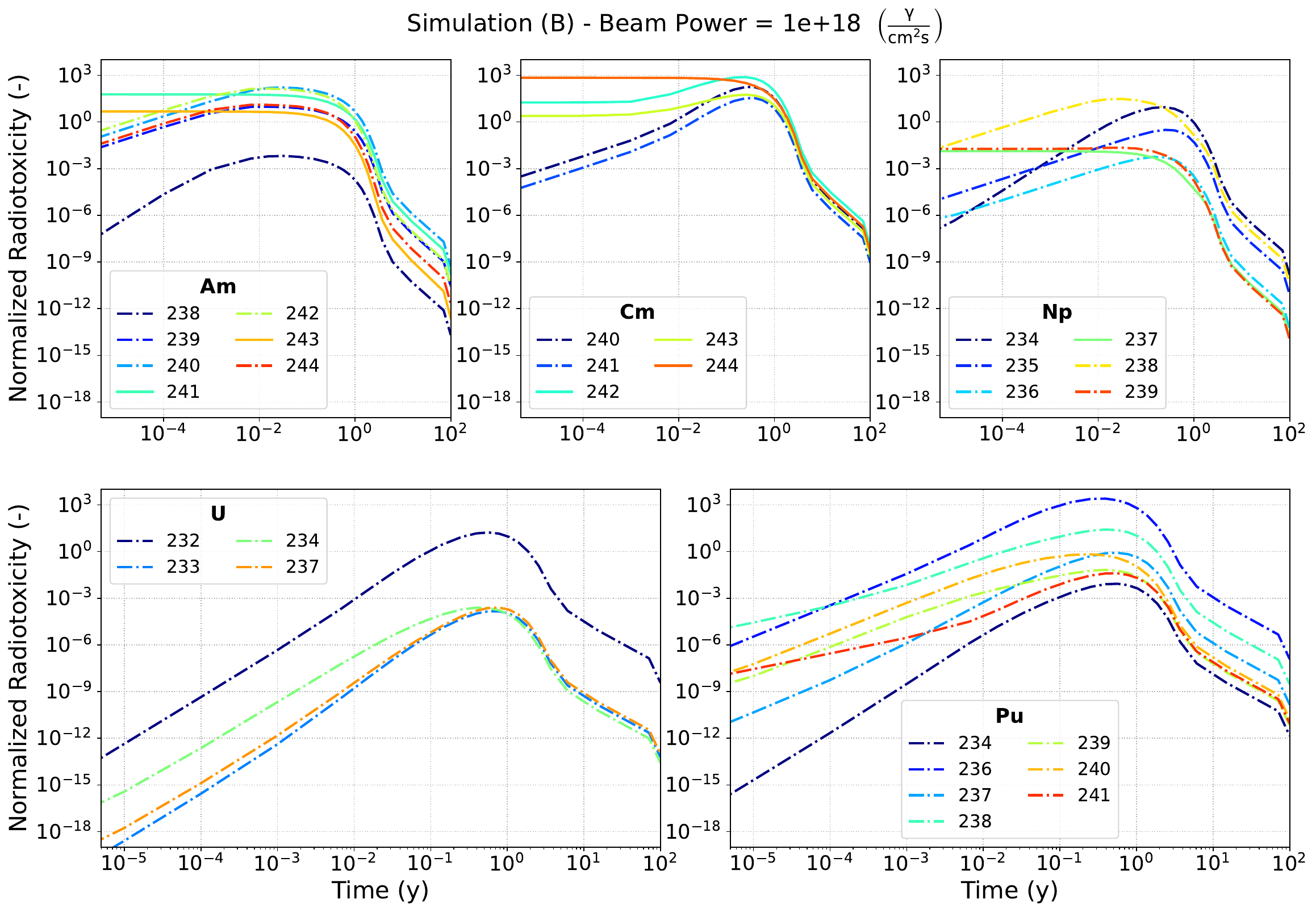}
    \caption{Radiotoxicity from minor actinides concentration in Simulation (B), considering a volume average flux of 1E+18 $\mathrm{\frac{\gamma}{cm^2 s}}$. Solid lines represent the nuclides considered in the initial condition, whereas dotted lines are the by-product.}
    \label{fig:radiotoxicity_isotopes_B}
\end{figure}

\begin{figure}
    \centering
    \includegraphics[width=1\linewidth]{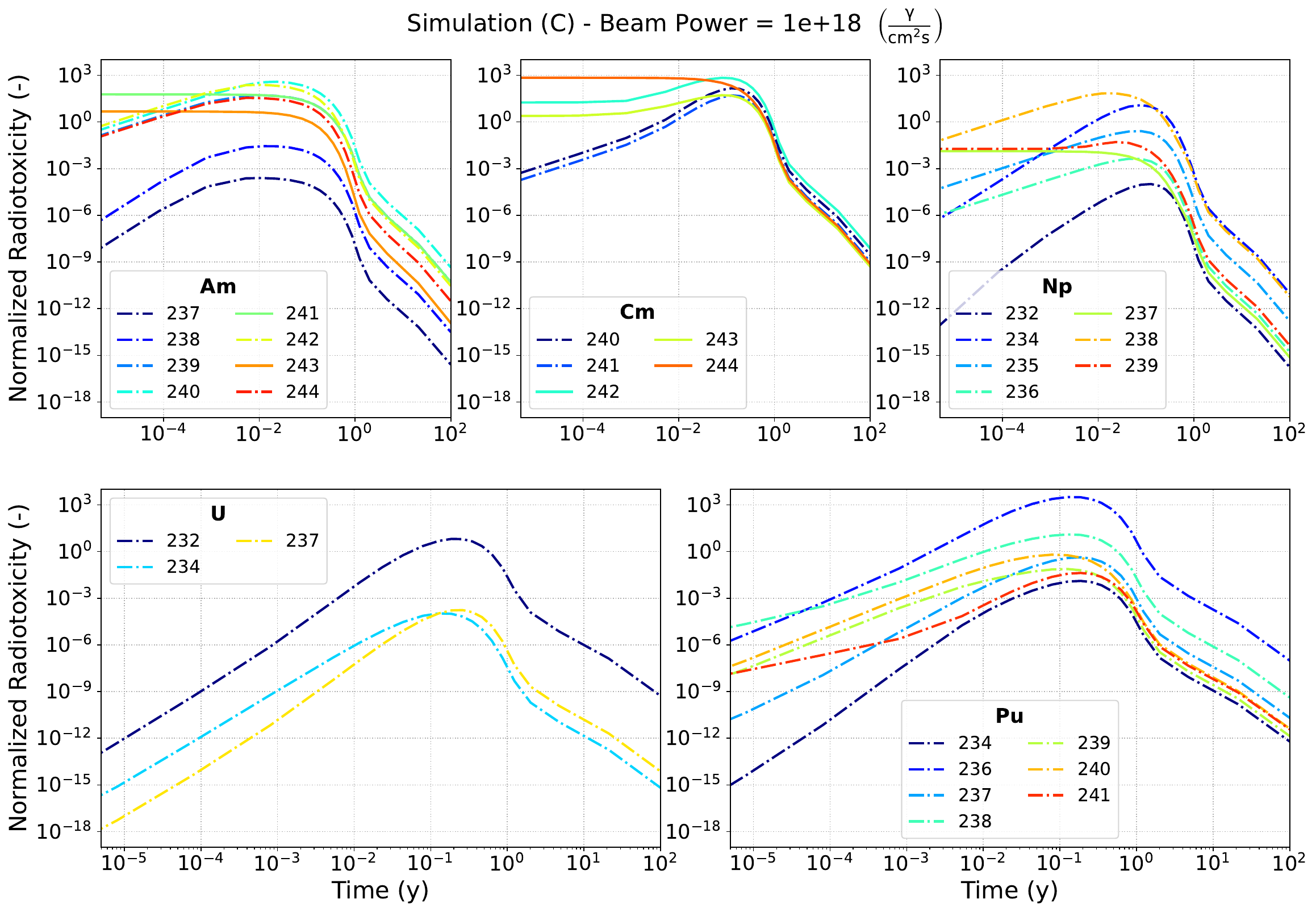}
    \caption{Radiotoxicity from minor actinides concentration in Simulation (C), considering a volume average flux of 1E+18 $\mathrm{\frac{\gamma}{cm^2 s}}$. Solid lines represent the nuclides considered in the initial condition, whereas dotted lines are the by-product.}
    \label{fig:radiotoxicity_isotopes_C}
\end{figure}

The previous section presented the time evolution of nuclides in the system, highlighting their decrease over the selected time window. However, a quantitative indicator is needed to assess the effectiveness of the treatment. As explained in Section~~\ref{sec:radiotoxicity_met}, radiotoxicity serves this purpose. 
Figures~\ref{fig:radiotoxicity_isotopes_B} and ~\ref{fig:radiotoxicity_isotopes_C} shows the time evolution of the concentrations weighted by their damage coefficient, therefore evaluating the contribution of each single isotope to the total radiotoxicity. This offers a complementary view on what is happening on the system: as previously shown, some by-product nuclides are present in traces, but their impact on the radiotoxicity is not negligible (e.g., $^{236}$Pu or $^{240}$Am).
By linearly combining the concentrations with the damage coefficients as given in Equations~\eqref{eqn:radiotoxicity_system} and \eqref{eqn:radiotoxicity_free} and then normalizing to the natural uranium radiotoxicity estimated in Equation~\eqref{eqn:radiotoxicity_natural}, it is possible to calculate when the threshold of natural radiotoxicity is crossed for the proposed scenarios.

\begin{figure}
    \centering
    \includegraphics[width=1\linewidth]{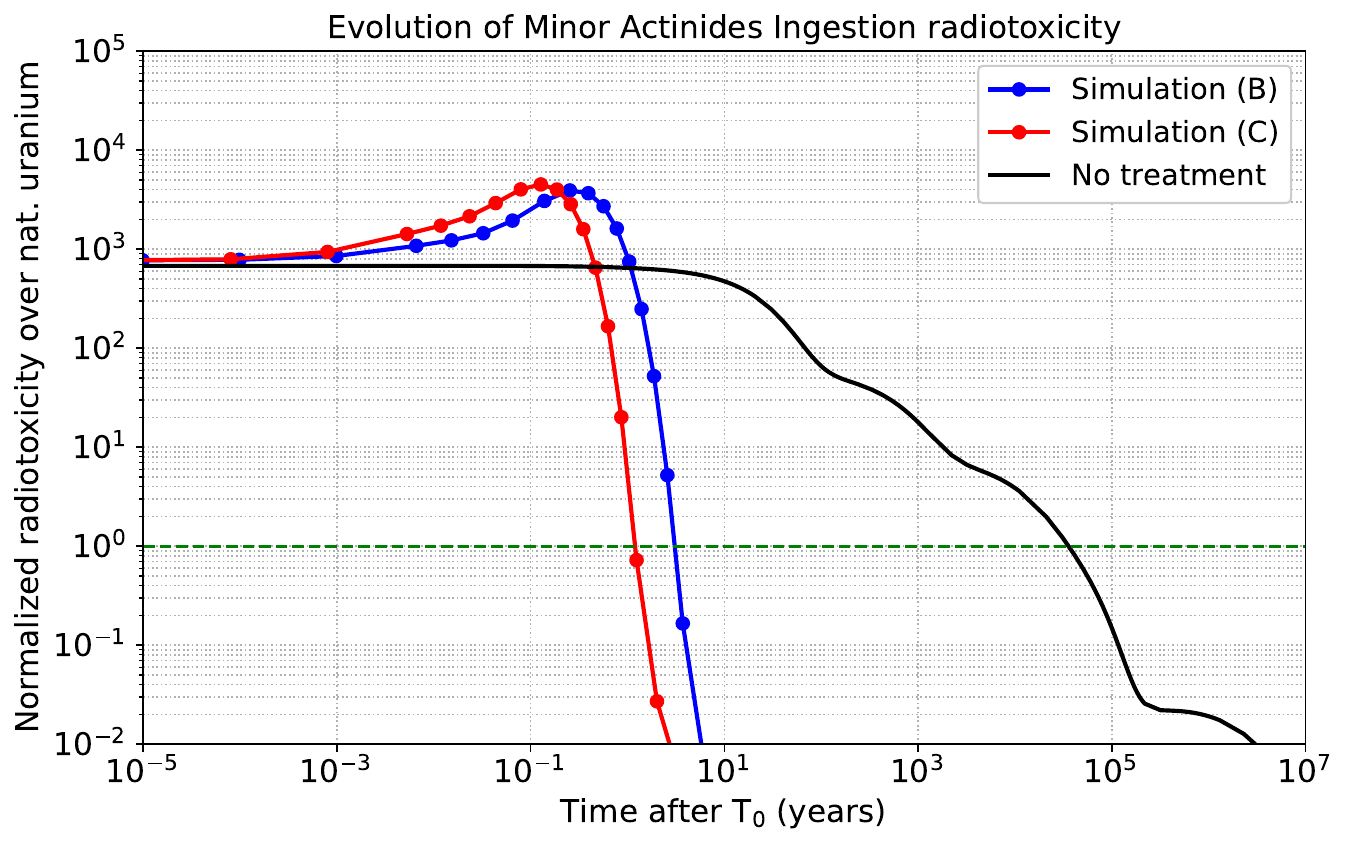}
    \caption{Normalized radiotoxicity evolution due to minor actinides in cases (B) and (C) when the sample is subjected to a flux of 1E+18 $\mathrm{\frac{\gamma}{cm^2 s}}$, compared with decay evolution.  The start time T$_0$ is a reference value that considers the time that the fuel stayed in the decay pool.}
    \label{fig:radiotoxicity}
\end{figure}

\begin{figure}
    \centering
    \includegraphics[width=1\linewidth]{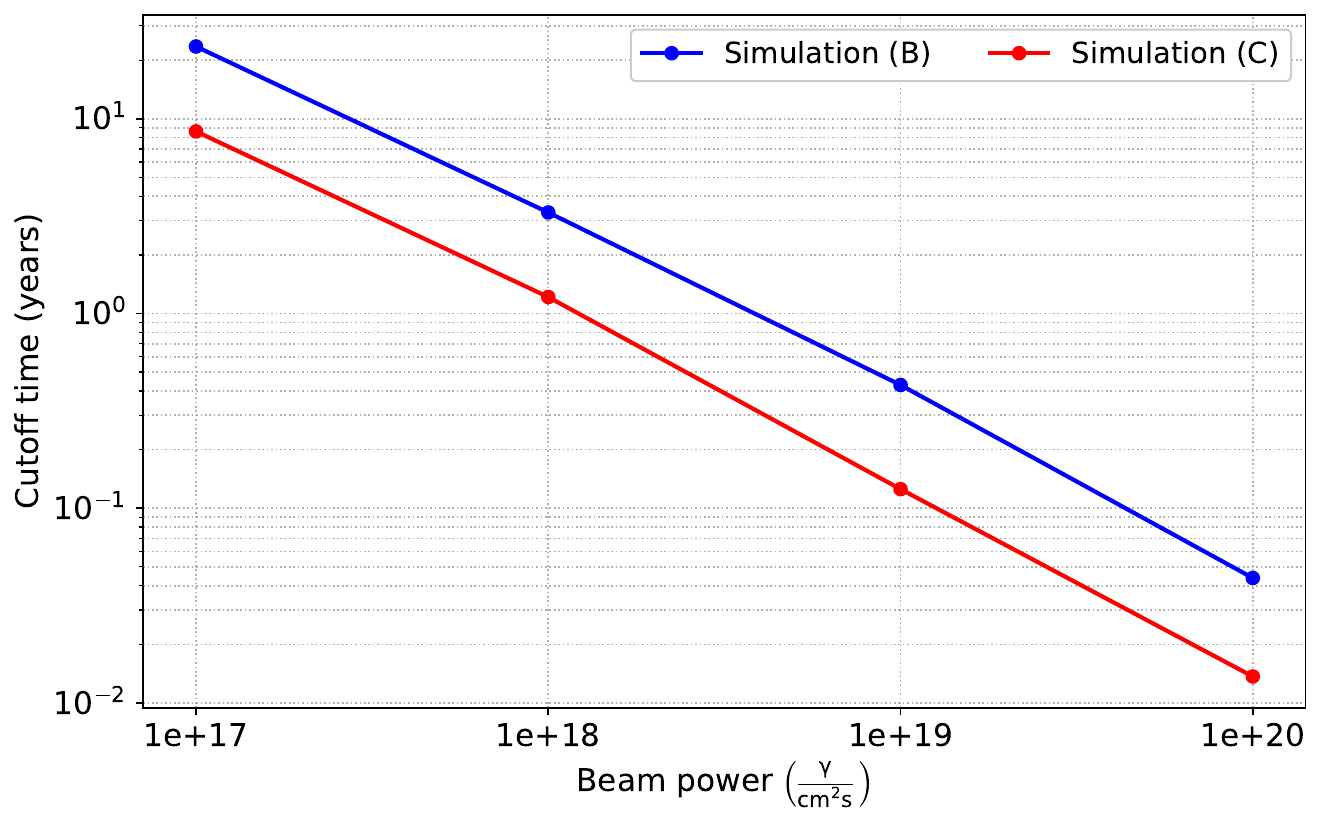}
    \caption{Time required for bringing the radiotoxicity of the initial spent fuel sample below the natural uranium level depending on the volume averaged flux.}
    \label{fig:cutoff_time}
\end{figure}

Figure~\ref{fig:radiotoxicity} shows the normalized radiotoxicity evolution due to the minor actinide content when the irradiation beam is given by Bremmstrahlung photons (i.e., Simulation B) or by LCB photons (i.e., Simulation C). Both cases reveal a similar trend: the initial radiotoxicity tends to increase up to a factor 5 due to the photon transmutations, leading the system to increase the radiotoxic nuclides content. Then, a rapid decrease is observed due to further transmutations to less radiotoxic species, as well as due to fission disintegrations. Simulation (C) appears to be more effective in terms of timing with respect to Simulation (B), crossing sooner the threshold. The shown radiotoxicity regards the MAs content only, not considering the transmutations that may occur to LLFPS, eventually increasing the system radiotoxicity.
The time in which the radiotoxicity curve crosses the value of 1, is called \textit{cut-off} time. Both curves are compared with the radiotoxicity trend in case of no SNF treatment. It is worth noticing that the evolution in all the three cases starts after two years the fuel was stored in a decay pool.\\
As mentioned in the previous subsection, the effectiveness of the concentration reduction varies depending on the averaged flux level. This is true also for radiotoxicity purposes, being the cut-off time dependent on the power. Figure~\ref{fig:cutoff_time} shows the time in which the threshold is crossed for the considered averaged flux levels in the two scenarios. As the flux increases, as the cut-off time decreases, with a linear trend, reaching $\sim$30 days in the best case scenario (Simulation (C) at 1E+20 $\mathrm{\frac{\gamma}{cm^2 s}}$).

\section{Conclusions}
\label{sec:conclusions}

This paper investigated the methodological approach of adopting photon-guided irradiation to achieve a reduction in minor actinides content, with a focus on the physical phenomenon and interaction in the spent fuel. The industrial feasibility (i.e., the cost estimation to maintain the considered fluxes) was not taken into account in this work. The presented methodology has been applied to spent fuel simulated from a SMR technology, specifically the NuScale design, but it remains general for every reactor concept. The analysis was conducted following a step-by-step approach, taking care to consider only the minor actinides present in the total spent fuel content. Three separate simulations were conducted, each with a different irradiation geometry: electron irradiation using a tantalum converter (A), direct conversion on the actinides (B), and photon irradiation from a laser combined beam (C).
The Bateman equations were solved for 53 nuclides with a Forward Euler scheme, accounting for their mutual decay chain as well as reactions resulting from interactions with photons and neutrons' transmutations and fissions. Several initial volume averaged fluxes have been considered, spanning from 1E+17 $\mathrm{\frac{\gamma}{cm^2 s}}$ to 1E+20 $\mathrm{\frac{\gamma}{cm^2 s}}$, chosen accordingly to literature reference. Through the adoption of damage coefficients for the nuclides involved in the analysis, the ingestion radiotoxicity was calculated and tracked over time, allowing for comparison with natural uranium radiotoxicity. Depending on the flux level employed, the cut-off time for the radiotoxicity of the spent fuel pellet decreases, following a linear trend with power, effectively burning $\sim $ 130 grams of Minor Actinides present in the spent fuel in approximately 30 days when a volume averaged flux of 1E+20 $\mathrm{\frac{\gamma}{cm^2 s}}$ is used in simulation (C).\\
The technologies examined, including LINAC, electron irradiation, and laser-Compton backscattering irradiation, were compared at the same flux level, showing the best performances of the quasi monoenergetic flux shape in the LCB case. However, all methods effectively utilized photonuclear reactions, suggesting to potentially become a viable alternatives for the transmutation of minor actinides if such fluxes were maintained in the system. In this scenario, the time required for these products to decrease to radiotoxicity levels comparable to natural uranium is highly reduced by several orders of magnitude.\\
While there are opportunities for improvement from a technological perspective, as higher flux intensities correlate with shorter irradiation times, the current analysis suggests that the LCB technology would be the most effective in terms of performance, being able to maximize the transmutation capabilities with the considered flux level. Overall, these findings underscore the potential of this method in addressing the challenges associated with nuclear waste treatment.\\
The fission product chains, along with their transmutation performances will be faced in future works, helping in the correct estimation of the system radiotoxicity reduction. Then, a cost estimation of the technology discussed will be essential to assess the competitiveness of this strategy for managing the spent fuel treatment, with the front-end power consumption of the accelerator included in this evaluation. In this framework, it is crucial to explore the potential of recovering energy produced during this process, to reduce the overall operational costs. To do so, some innovative irradiation configurations will be investigated, including the use of liquid-phase spent-fuel, enhancing the heat recovery with the natural circulation phenomenon.

\section*{List of Symbols}
\textbf{Acronyms}

\begin{longtable}{rl}
\textbf{ADS}    & Accelerator Driven System \\[5pt]
\textbf{BA}    & Burnable Absorber \\[5pt]
\textbf{FP}    & Fission Product \\[5pt]
\textbf{GDR}    & Giant Dipole Resonance \\[5pt]
\textbf{IAEA}    & International Atomic Energy Agency \\[5pt]
\textbf{LCB}    & Laser Compton Backscattering \\[5pt]
\textbf{LLFP}    & Long Lived Fission Product \\[5pt]
\textbf{LWR}    & Light Water Reactor \\[5pt]
\textbf{MA}    & Minor Actinide \\[5pt]
\textbf{NPP}    & Nuclear Power Plant \\[5pt]
\textbf{P\&T}    & Partitioning and Transmutation \\[5pt]
\textbf{PWR}    & Pressurized Water Reactor \\[5pt]
\textbf{SMR}    & Small Modular Reactor \\[5pt]
\textbf{SNF}    & Spent Nuclear Fuel \\[5pt]

\end{longtable}
\addtocounter{table}{-1}

\textbf{Greek Symbols}
\begin{longtable}{rl}
{$\gamma$} & {Photon} \\
{$\lambda$} & {Decay constant} \\
{$\omega$} & {Frequency} \\
{$\sigma$} & {Microscopic cross section} \\
{$\theta$} & {Scattering angle} \\
{$\Phi$} & {Energy flux} \\

\end{longtable}
\addtocounter{table}{-1}

\textbf{Latin symbols}
\begin{longtable}{rl}
{$E$} & {Energy} \\
{$F_s$} & {Tuning factor} \\
{$P$} & {Volume averaged flux} \\
{$R$} & {Radiotoxicity} \\
{$N$} & {Atomic concentration} \\
{$V$} & {Volume} \\
{$f$} & {Fission} \\
{$h$} & {Planck constant} \\
{$n$} & {Neutron} \\
{$t$} & {Time} \\
{$y$} & {Dose coefficient} \\

\end{longtable}
\addtocounter{table}{-1}

\newpage
\bibliography{bibliography.bib}

\begin{thebibliography}{10}
\expandafter\ifx\csname url\endcsname\relax
  \def\url#1{\texttt{#1}}\fi
\expandafter\ifx\csname urlprefix\endcsname\relax\def\urlprefix{URL }\fi
\expandafter\ifx\csname href\endcsname\relax
  \def\href#1#2{#2} \def\path#1{#1}\fi

\bibitem{IAEA:2023}
IAEA, \href{https://www.iaea.org/publications/15197/global-status-of-decommissioning-of-nuclear-installations}{Global Status of Decommissioning of Nuclear Installations}, no. NW-T-2.16 in Nuclear Energy Series, International Atomic Energy Agency, Vienna, 2023.
\newline\urlprefix\url{https://www.iaea.org/publications/15197/global-status-of-decommissioning-of-nuclear-installations}

\bibitem{MCFARLANE2004351}
H.~F. McFarlane, \href{https://www.sciencedirect.com/science/article/pii/B012176480X002989}{Nuclear fuel reprocessing}, in: C.~J. Cleveland (Ed.), Encyclopedia of Energy, Elsevier, New York, 2004, pp. 351--364.
\newblock \href {https://doi.org/https://doi.org/10.1016/B0-12-176480-X/00298-9} {\path{doi:https://doi.org/10.1016/B0-12-176480-X/00298-9}}.
\newline\urlprefix\url{https://www.sciencedirect.com/science/article/pii/B012176480X002989}

\bibitem{ORNL:2024}
D.~Hartanto, G.~Radulescu, R.~Bostelmann, R.~Elzohery, W.~Wieselquist, Scale demonstration for sodium-cooled fast reactor fuel cycle analysis, Tech. rep., Oak Ridge National Laboratory (ORNL), Oak Ridge, TN (United States) (2024).

\bibitem{stanford:2022}
{Stanford News}, \href{https://news.stanford.edu/stories/2022/05/small-modular-reactors-produce-high-levels-nuclear-waste}{Stanford-led research finds small modular reactors will exacerbate challenges of highly radioactive nuclear waste}, accessed: 2024-09-18 (2022).
\newline\urlprefix\url{https://news.stanford.edu/stories/2022/05/small-modular-reactors-produce-high-levels-nuclear-waste}

\bibitem{OLANDER20014490}
D.~Olander, \href{https://www.sciencedirect.com/science/article/pii/B0080431526007877}{Light water reactor fuel design and performance}, in: K.~J. Buschow, R.~W. Cahn, M.~C. Flemings, B.~Ilschner, E.~J. Kramer, S.~Mahajan, P.~Veyssière (Eds.), Encyclopedia of Materials: Science and Technology, Elsevier, Oxford, 2001, pp. 4490--4504.
\newblock \href {https://doi.org/https://doi.org/10.1016/B0-08-043152-6/00787-7} {\path{doi:https://doi.org/10.1016/B0-08-043152-6/00787-7}}.
\newline\urlprefix\url{https://www.sciencedirect.com/science/article/pii/B0080431526007877}

\bibitem{Office:2018}
U.~Office, \href{https://books.google.it/books?id=8tNgswEACAAJ}{Nuclear Fuel Cycle Options: Doe Needs to Enhance Planning for Technology Assessment and Collaboration with Industry and Other Countries}, CreateSpace Independent Publishing Platform, 2018.
\newline\urlprefix\url{https://books.google.it/books?id=8tNgswEACAAJ}

\bibitem{benke2023finland}
E.~Benke, Finland's plan to bury spent nuclear fuel for 100,000 years, \url{https://www.bbc.com/news/science-environment-66304667}, accessed: 2024-08-17 (2023).

\bibitem{onkalo2023}
Onkalo nuclear waste disposal facility, olkiluoto, finland, \url{https://www.nsenergybusiness.com/projects/onkalo-nuclear-waste-disposal-facility/}, retrieved 2024-08-17 (2023).

\bibitem{Rubbia:2001}
C.~Rubbia, H.~Ait~Abderrahim, M.~Björnberg, B.~Carluec, G.~Gherardi, E.~Romero, W.~Gudowski, G.~Heusener, H.~Leeb, W.~von Lensa, G.~Locatelli, J.~Magill, J.~Martinez-Val, S.~Monti, A.~Mueller, M.~Napolitano, A.~Pérez-Navarro, M.~Salvatores, J.~Soares, J.~Thomas, The european roadmap for developing ads for nuclear waste incineration, Tech. rep., ENEA (04 2001).

\bibitem{OECD:2006}
OECD, Physics and safety of transmutation systems: A status report, OECD Papers 6 (2006) 13--13.

\bibitem{Kooyman:2021}
T.~Kooyman, \href{https://www.sciencedirect.com/science/article/pii/S0306454921001158}{Current state of partitioning and transmutation studies for advanced nuclear fuel cycles}, Annals of Nuclear Energy 157 (2021) 108239.
\newblock \href {https://doi.org/https://doi.org/10.1016/j.anucene.2021.108239} {\path{doi:https://doi.org/10.1016/j.anucene.2021.108239}}.
\newline\urlprefix\url{https://www.sciencedirect.com/science/article/pii/S0306454921001158}

\bibitem{ALFRED:2013}
M.~Frogheri, A.~Alemberti, L.~Mansani, The lead fast reactor: Demonstrator (alfred) and elfr design, in: nternational Conference on Fast Reactors and Related Fuel Cycles: Safe Technologies and Sustainable Scenarios, 2013, p.~15.

\bibitem{MSFR:concept}
M.~Allibert, S.~Delpech, D.~Gerardin, D.~Heuer, A.~Laureau, E.~Merle, \href{https://www.sciencedirect.com/science/article/pii/B9780128205884000050}{Chapter 7 - homogeneous molten salt reactors (msrs): The molten salt fast reactor (msfr) concept}, in: I.~L. Pioro (Ed.), Handbook of Generation IV Nuclear Reactors (Second Edition), second edition Edition, Woodhead Publishing Series in Energy, Woodhead Publishing, 2023, pp. 231--257.
\newblock \href {https://doi.org/https://doi.org/10.1016/B978-0-12-820588-4.00005-0} {\path{doi:https://doi.org/10.1016/B978-0-12-820588-4.00005-0}}.
\newline\urlprefix\url{https://www.sciencedirect.com/science/article/pii/B9780128205884000050}

\bibitem{Ridikas:2002}
D.~Ridikas, H.~Safa, M.-L. Mauborgne, Conceptual study of neutron irradiator driven by electron accelerator, in: 7th Information Exchange Meeting on Actinide and Fission Product P\&T (NEA/OCDE), 2002, pp.~--.

\bibitem{Chen:2008}
C.~Jin-Gen, X.~Wang, W.~Hong-Wei, G.~Wei, M.~Yu-Gang, C.~Xiang-Zhou, L.~Guang-Cheng, X.~Yi, P.~Qiang-Yan, Y.~Ren-Yong, X.~Jia-Qiang, Y.~Zhe, F.~Gong-Tao, S.~Wen-Qing, \href{https://dx.doi.org/10.1088/1674-1137/32/8/019}{Transmutation of nuclear wastes using photonuclear reactions triggered by compton backscattering photons at the shanghai laser electron gamma source*}, Chinese Physics C 32~(8) (2008) 677.
\newblock \href {https://doi.org/10.1088/1674-1137/32/8/019} {\path{doi:10.1088/1674-1137/32/8/019}}.
\newline\urlprefix\url{https://dx.doi.org/10.1088/1674-1137/32/8/019}

\bibitem{Ledingham:2010}
K.~Ledingham, W.~Galster, Laser-driven particle and photon beams and some applications, New Journal of Physics 12~(April) (Apr. 2010).
\newblock \href {https://doi.org/10.1088/1367-2630/12/4/045005} {\path{doi:10.1088/1367-2630/12/4/045005}}.

\bibitem{Wang:2017}
X.~Wang, Z.~Xu, L.~Wen, H.~Lu, Z.~Zhu, X.~Yan, Transmutation prospect of long-lived nuclear waste induced by high-charge electron beam from laser plasma accelerator, Physics of Plasmas 24 (05 2017).
\newblock \href {https://doi.org/10.1063/1.4998470} {\path{doi:10.1063/1.4998470}}.

\bibitem{Matsumoto:1988}
T.~Matsumoto, \href{https://www.sciencedirect.com/science/article/pii/0168900288906134}{Calculations of gamma ray incineration of 90sr and 137cs}, Nuclear Instruments and Methods in Physics Research Section A: Accelerators, Spectrometers, Detectors and Associated Equipment 268~(1) (1988) 234--243.
\newblock \href {https://doi.org/https://doi.org/10.1016/0168-9002(88)90613-4} {\path{doi:https://doi.org/10.1016/0168-9002(88)90613-4}}.
\newline\urlprefix\url{https://www.sciencedirect.com/science/article/pii/0168900288906134}

\bibitem{Wickert:2024}
C.~Wickert, G.~Tukharyan, J.~Yu, A.~Danagoulian, Z.~Hartwig, B.~Forget, Progress on transport and transmutation analysis of proton beam irradiating long-lived spent nuclear waste, in: Proceedings of the International Conference on Physics of Reactors (PHYSOR), American Nuclear Society, San Francisco, CA, 2024, pp. 55--62.
\newblock \href {https://doi.org/doi.org/10.13182/PHYSOR24-43724} {\path{doi:doi.org/10.13182/PHYSOR24-43724}}.

\bibitem{Diamond:1999}
W.~T. Diamond, \href{https://doi.org/10.1016/S0168-9002(99)00492-1}{A radioactive ion beam facility using photofission}, Nuclear Instruments and Methods in Physics Research Section A: Accelerators, Spectrometers, Detectors and Associated Equipment 432~(2-3) (1999) 471--482.
\newblock \href {https://doi.org/10.1016/S0168-9002(99)00492-1} {\path{doi:10.1016/S0168-9002(99)00492-1}}.
\newline\urlprefix\url{https://doi.org/10.1016/S0168-9002(99)00492-1}

\bibitem{Balabanski:2024znh}
D.~L. Balabanski, P.~Constantin, {80 years of experimental photo-fission research}, Eur. Phys. J. A 60~(2) (2024) 39.
\newblock \href {https://doi.org/10.1140/epja/s10050-024-01264-z} {\path{doi:10.1140/epja/s10050-024-01264-z}}.

\bibitem{NuScale}
E.~Diaz-Pescador, Y.~Bilodid, M.~Jobst, S.~Kliem, \href{https://www.sciencedirect.com/science/article/pii/S0029549324000116}{Nuscale-like smr model development and applied safety analyses with the code chain serpent-dyn3d-athlet}, Nuclear Engineering and Design 418 (2024) 112909.
\newblock \href {https://doi.org/https://doi.org/10.1016/j.nucengdes.2024.112909} {\path{doi:https://doi.org/10.1016/j.nucengdes.2024.112909}}.
\newline\urlprefix\url{https://www.sciencedirect.com/science/article/pii/S0029549324000116}

\bibitem{WILLAT:2023}
R.~Willat, P.~Deng, W.~Yang, A feasibility study of photonuclear transmutation of long-lived fission products without isotopic separation, Trans. Am. Nucl. Soc. 128 (2023) 127.

\bibitem{Bengt:1972}
B.~Forkman, B.~Schrøder, \href{https://dx.doi.org/10.1088/0031-8949/5/3/001}{A review of intermediate energy photofission}, Physica Scripta 5~(3) (1972) 105.
\newblock \href {https://doi.org/10.1088/0031-8949/5/3/001} {\path{doi:10.1088/0031-8949/5/3/001}}.
\newline\urlprefix\url{https://dx.doi.org/10.1088/0031-8949/5/3/001}

\bibitem{JENDL5}
I.~Osamu, I.~Nobuyuki, K.~Satoshi, M.~Futoshi, N.~Shinsuke, A.~Yutaka, T.~Kohsuke, O.~Shin, I.~Chikako, Y.~Tadashi, C.~Satoshi, O.~Naohiko, S.~Jean-Christophe, I.~Hiroki, Y.~Kazuyoshi, N.~Yasunobu, T.~Kenichi, K.~Chikara, M.~Norihiro, Y.~Kenji, T.~Hiroshi, O.~Akitom, F.~Masahiro, O.~Shoichiro, C.~Go, S.~Satoshi, O.~Masayuki, K.~Saerom, \href{https://doi.org/10.1080/00223131.2022.2141903}{Japanese evaluated nuclear data library version 5: Jendl-5}, Journal of Nuclear Science and Technology 60~(1) (2023) 1--60.
\newblock \href {http://arxiv.org/abs/https://doi.org/10.1080/00223131.2022.2141903} {\path{arXiv:https://doi.org/10.1080/00223131.2022.2141903}}, \href {https://doi.org/10.1080/00223131.2022.2141903} {\path{doi:10.1080/00223131.2022.2141903}}.
\newline\urlprefix\url{https://doi.org/10.1080/00223131.2022.2141903}

\bibitem{TENDL}
A.~Koning, D.~Rochman, J.-C. Sublet, N.~Dzysiuk, M.~Fleming, S.~{van der Marck}, \href{https://www.sciencedirect.com/science/article/pii/S009037521930002X}{Tendl: Complete nuclear data library for innovative nuclear science and technology}, Nuclear Data Sheets 155 (2019) 1--55, special Issue on Nuclear Reaction Data.
\newblock \href {https://doi.org/https://doi.org/10.1016/j.nds.2019.01.002} {\path{doi:https://doi.org/10.1016/j.nds.2019.01.002}}.
\newline\urlprefix\url{https://www.sciencedirect.com/science/article/pii/S009037521930002X}

\bibitem{ENDF8}
D.~Brown, M.~Chadwick, R.~Capote, A.~Kahler, A.~Trkov, M.~Herman, A.~Sonzogni, Y.~Danon, A.~Carlson, M.~Dunn, D.~Smith, G.~Hale, G.~Arbanas, R.~Arcilla, C.~Bates, B.~Beck, B.~Becker, F.~Brown, R.~Casperson, J.~Conlin, D.~Cullen, M.-A. Descalle, R.~Firestone, T.~Gaines, K.~Guber, A.~Hawari, J.~Holmes, T.~Johnson, T.~Kawano, B.~Kiedrowski, A.~Koning, S.~Kopecky, L.~Leal, J.~Lestone, C.~Lubitz, J.~M. Dami{\'a}n, C.~Mattoon, E.~McCutchan, S.~Mughabghab, P.~Navratil, D.~Neudecker, G.~Nobre, G.~Noguere, M.~Paris, M.~Pigni, A.~Plompen, B.~Pritychenko, V.~Pronyaev, D.~Roubtsov, D.~Rochman, P.~Romano, P.~Schillebeeckx, S.~Simakov, M.~Sin, I.~Sirakov, B.~Sleaford, V.~Sobes, E.~Soukhovitskii, I.~Stetcu, P.~Talou, I.~Thompson, S.~van~der Marck, L.~Welser-Sherrill, D.~Wiarda, M.~White, J.~Wormald, R.~Wright, M.~Zerkle, G.~\v{Z}erovnik, Y.~Zhu, \href{https://www.sciencedirect.com/science/article/pii/S0090375218300206}{{ENDF/B-VIII.0}: The {8$^{th}$} major release of the nuclear reaction data library with {CIELO}-project
  cross sections, new standards and thermal scattering data}, Nuclear Data Sheets 148 (2018) 1 -- 142, special Issue on Nuclear Reaction Data.
\newblock \href {https://doi.org/https://doi.org/10.1016/j.nds.2018.02.001} {\path{doi:https://doi.org/10.1016/j.nds.2018.02.001}}.
\newline\urlprefix\url{https://www.sciencedirect.com/science/article/pii/S0090375218300206}

\bibitem{ECKERMAN20121}
K.~Eckerman, J.~Harrison, H.-G. Menzel, C.~Clement, \href{https://www.sciencedirect.com/science/article/pii/S014664531200053X}{Icrp publication 119: Compendium of dose coefficients based on icrp publication 60}, Annals of the ICRP 41 (2012) 1--130, iCRP PUBLICATION 119:.
\newblock \href {https://doi.org/https://doi.org/10.1016/j.icrp.2012.06.038} {\path{doi:https://doi.org/10.1016/j.icrp.2012.06.038}}.
\newline\urlprefix\url{https://www.sciencedirect.com/science/article/pii/S014664531200053X}

\bibitem{endfviii}
D.~Brown, M.~Chadwick, R.~Capote, A.~Kahler, A.~Trkov, M.~Herman, A.~Sonzogni, Y.~Danon, A.~Carlson, M.~Dunn, D.~Smith, G.~Hale, G.~Arbanas, R.~Arcilla, C.~Bates, B.~Beck, B.~Becker, F.~Brown, R.~Casperson, J.~Conlin, D.~Cullen, M.-A. Descalle, R.~Firestone, T.~Gaines, K.~Guber, A.~Hawari, J.~Holmes, T.~Johnson, T.~Kawano, B.~Kiedrowski, A.~Koning, S.~Kopecky, L.~Leal, J.~Lestone, C.~Lubitz, J.~M. Dami{\'a}n, C.~Mattoon, E.~McCutchan, S.~Mughabghab, P.~Navratil, D.~Neudecker, G.~Nobre, G.~Noguere, M.~Paris, M.~Pigni, A.~Plompen, B.~Pritychenko, V.~Pronyaev, D.~Roubtsov, D.~Rochman, P.~Romano, P.~Schillebeeckx, S.~Simakov, M.~Sin, I.~Sirakov, B.~Sleaford, V.~Sobes, E.~Soukhovitskii, I.~Stetcu, P.~Talou, I.~Thompson, S.~van~der Marck, L.~Welser-Sherrill, D.~Wiarda, M.~White, J.~Wormald, R.~Wright, M.~Zerkle, G.~\v{Z}erovnik, Y.~Zhu, \href{https://www.sciencedirect.com/science/article/pii/S0090375218300206}{{ENDF/B-VIII.0}: The {8$^{th}$} major release of the nuclear reaction data library with {CIELO}-project
  cross sections, new standards and thermal scattering data}, Nuclear Data Sheets 148 (2018) 1 -- 142, special Issue on Nuclear Reaction Data.
\newblock \href {https://doi.org/https://doi.org/10.1016/j.nds.2018.02.001} {\path{doi:https://doi.org/10.1016/j.nds.2018.02.001}}.
\newline\urlprefix\url{https://www.sciencedirect.com/science/article/pii/S0090375218300206}

\bibitem{Wigeland:2015}
R.~Wigeland, T.~Taiwo, J.~Gehin, R.~Jubin, M.~Todosow, Nuclear fuel cycle evaluation and screening findings on partitioning and transmutation (giu 2015).

\bibitem{SERPENT}
J.~Leppänen, M.~Pusa, T.~Viitanen, V.~Valtavirta, T.~Kaltiaisenaho, \href{https://www.sciencedirect.com/science/article/pii/S0306454914004095}{The serpent monte carlo code: Status, development and applications in 2013}, Annals of Nuclear Energy 82 (2015) 142--150.
\newblock \href {https://doi.org/https://doi.org/10.1016/j.anucene.2014.08.024} {\path{doi:https://doi.org/10.1016/j.anucene.2014.08.024}}.
\newline\urlprefix\url{https://www.sciencedirect.com/science/article/pii/S0306454914004095}

\bibitem{MonteCarlo_SIE_stability}
J.~Dufek, D.~Kotlyar, E.~Shwageraus, \href{[https://www.sciencedirect.com/science/article/pii/S0306454913002703](https://www.sciencedirect.com/science/article/pii/S0306454913002703)}{{The stochastic implicit Euler method – A stable coupling scheme for Monte Carlo burnup calculations}}, Annals of Nuclear Energy 60 (2013).
\newblock \href {https://doi.org/[https://doi.org/10.1016/j.anucene.2013.05.015](https://doi.org/10.1016/j.anucene.2013.05.015)} {\path{doi:[https://doi.org/10.1016/j.anucene.2013.05.015](https://doi.org/10.1016/j.anucene.2013.05.015)}}.
\newline\urlprefix\url{[https://www.sciencedirect.com/science/article/pii/S0306454913002703](https://www.sciencedirect.com/science/article/pii/S0306454913002703)}

\bibitem{Bairiot:1992}
H.~Bairiot, P.~Deramaix, Mox fuel development: yesterday, today and tomorrow, Journal of Nuclear Materials 188 (1992) 10--18.
\newblock \href {https://doi.org/10.1016/0022-3115(92)90448-T} {\path{doi:10.1016/0022-3115(92)90448-T}}.

\bibitem{BATTISTONI201510}
G.~Battistoni, T.~Boehlen, F.~Cerutti, P.~W. Chin, L.~S. Esposito, A.~Fassò, A.~Ferrari, A.~Lechner, A.~Empl, A.~Mairani, A.~Mereghetti, P.~G. Ortega, J.~Ranft, S.~Roesler, P.~R. Sala, V.~Vlachoudis, G.~Smirnov, \href{https://www.sciencedirect.com/science/article/pii/S0306454914005878}{Overview of the fluka code}, Annals of Nuclear Energy 82 (2015) 10--18, joint International Conference on Supercomputing in Nuclear Applications and Monte Carlo 2013, SNA + MC 2013. Pluri- and Trans-disciplinarity, Towards New Modeling and Numerical Simulation Paradigms.
\newblock \href {https://doi.org/https://doi.org/10.1016/j.anucene.2014.11.007} {\path{doi:https://doi.org/10.1016/j.anucene.2014.11.007}}.
\newline\urlprefix\url{https://www.sciencedirect.com/science/article/pii/S0306454914005878}

\bibitem{Oleinikov:2023}
E.~Oleinikov, I.~Pylypchynets, O.~Parlag, Simulation of the concomitant nuclear reactions contribution to the actinide photofission on the m-30 microtron at 17.5 mev bremsstrahlung energy, New Journal of Nuclear and Particle Physics 13~(2) (2023).
\newblock \href {https://doi.org/10.1088/1367-2630/12/4/045005} {\path{doi:10.1088/1367-2630/12/4/045005}}.

\bibitem{2020SciPy}
P.~Virtanen, R.~Gommers, T.~E. Oliphant, M.~Haberland, T.~Reddy, D.~Cournapeau, E.~Burovski, P.~Peterson, W.~Weckesser, J.~Bright, S.~J. {van der Walt}, M.~Brett, J.~Wilson, K.~J. Millman, N.~Mayorov, A.~R.~J. Nelson, E.~Jones, R.~Kern, E.~Larson, C.~J. Carey, {\.I}.~Polat, Y.~Feng, E.~W. Moore, J.~{VanderPlas}, D.~Laxalde, J.~Perktold, R.~Cimrman, I.~Henriksen, E.~A. Quintero, C.~R. Harris, A.~M. Archibald, A.~H. Ribeiro, F.~Pedregosa, P.~{van Mulbregt}, {SciPy 1.0 Contributors}, {{SciPy} 1.0: Fundamental Algorithms for Scientific Computing in Python}, Nature Methods 17 (2020) 261--272.
\newblock \href {https://doi.org/10.1038/s41592-019-0686-2} {\path{doi:10.1038/s41592-019-0686-2}}.

\end{thebibliography}

\end{document}